\pdfoutput=1
\documentclass[12pt,a4paper]{article}
\usepackage{ifthen} % for conditional statements
\newboolean{pdflatex}
\setboolean{pdflatex}{true} % False for eps figures 

\newboolean{articletitles}
\setboolean{articletitles}{true} % False removes titles in references

\newboolean{uprightparticles}
\setboolean{uprightparticles}{false} %True for upright particle symbols

\def\paperauthors{LHCb collaboration} % Leave as is for PAPER, CONF and FIGURE
\def\paperasciititle{Improved measurement of eta/eta' mixing in B -> J/psi eta(') decays} % Set ASCII title here !! MAKE sure it's only ASCII characters !! 
\def\papertitle{Improved measurement of \etaz/\etapr mixing in $\Bds\to\jpsi\etazpr$ decays} % Latex formatted title
\def\paperkeywords{{High Energy Physics}, {LHCb}} % Comma separated list
\def\papercopyright{\the\year\ CERN for the benefit of the LHCb collaboration} % new since 9/Apr/2018
\def\paperlicence{CC BY 4.0 licence}
\def\paperlicenceurl{https://creativecommons.org/licenses/by/4.0/}

\newif\ifEnableSectionTOCLinks
\EnableSectionTOCLinksfalse % deactivated

\usepackage[top=1in, bottom=1.25in, left=1in, right=1in]{geometry}

\usepackage{microtype}
\usepackage{lineno}  % for line numbering during review
\usepackage{xspace} % To avoid problems with missing or double spaces after
\usepackage{caption} %these three command get the figure and table captions automatically small

\usepackage{graphicx}  % to include figures (can also use other packages)
\usepackage{color}
\usepackage{colortbl}
\graphicspath{{./figs/}} % Make Latex search fig subdir for figures
\usepackage{amsmath} % Adds a large collection of math symbols
\usepackage{amssymb}
\usepackage{amsfonts}
\usepackage{upgreek} % Adds in support for greek letters in roman typeset

\newcommand*\patchAmsMathEnvironmentForLineno[1]{%
\expandafter\let\csname old#1\expandafter\endcsname\csname #1\endcsname
\expandafter\let\csname oldend#1\expandafter\endcsname\csname
end#1\endcsname
 \renewenvironment{#1}%
   {\linenomath\csname old#1\endcsname}%
   {\csname oldend#1\endcsname\endlinenomath}%
}
\newcommand*\patchBothAmsMathEnvironmentsForLineno[1]{%
  \patchAmsMathEnvironmentForLineno{#1}%
  \patchAmsMathEnvironmentForLineno{#1*}%
}
\AtBeginDocument{%
\patchBothAmsMathEnvironmentsForLineno{equation}%
\patchBothAmsMathEnvironmentsForLineno{align}%
\patchBothAmsMathEnvironmentsForLineno{flalign}%
\patchBothAmsMathEnvironmentsForLineno{alignat}%
\patchBothAmsMathEnvironmentsForLineno{gather}%
\patchBothAmsMathEnvironmentsForLineno{multline}%
\patchBothAmsMathEnvironmentsForLineno{eqnarray}%
}

\usepackage[pdftex,
            pdfauthor={\paperauthors},
            pdftitle={\paperasciititle},
            pdfkeywords={\paperkeywords}]{hyperref}
\usepackage{hyperxmp}
\hypersetup{
    pdfcopyright={Copyright (C) \papercopyright},
    pdflicenseurl={\paperlicenceurl}
}
\usepackage[colorinlistoftodos,textsize=scriptsize]{todonotes}

\usepackage[bottom,flushmargin,hang,multiple]{footmisc}

\usepackage[all]{hypcap} % Internal hyperlinks to floats.

\usepackage{xspace}
\usepackage{upgreek}

\def\lhcb   {\mbox{LHCb}\xspace}

\def\MagUp {\mbox{\em Mag\kern -0.05em Up}\xspace}

\ifthenelse{\boolean{uprightparticles}}%
{
 
 \def\Pgamma      {\ensuremath{\upgamma}\xspace}

 \def\Peta        {\ensuremath{\upeta}\xspace}

 \def\Pmu         {\ensuremath{\upmu}\xspace}
 \def\Pnu         {\ensuremath{\upnu}\xspace}
 
 \def\Ppi         {\ensuremath{\uppi}\xspace}
 
 \def\Prho        {\ensuremath{\uprho}\xspace}

 \def\Pphi        {\ensuremath{\upphi}\xspace}

 \def\Ppsi        {\ensuremath{\uppsi}\xspace}

 \def\PDelta      {\ensuremath{\Delta}\xspace}
 \def\PXi         {\ensuremath{\Xi}\xspace}
 \def\PLambda     {\ensuremath{\Lambda}\xspace}
 \def\PSigma      {\ensuremath{\Sigma}\xspace}
 \def\POmega      {\ensuremath{\Omega}\xspace}
 \def\PUpsilon    {\ensuremath{\Upsilon}\xspace}
 \let\oldPi\Pi
 \def\PPi         {\ensuremath{\oldPi}\xspace}

 \def\PB      {\ensuremath{\mathrm{B}}\xspace}
 \def\PD      {\ensuremath{\mathrm{D}}\xspace}
 \def\PJ      {\ensuremath{\mathrm{J}}\xspace}
 \def\PK      {\ensuremath{\mathrm{K}}\xspace}
 \def\Pb      {\ensuremath{\mathrm{b}}\xspace}
 \def\Pc      {\ensuremath{\mathrm{c}}\xspace}
 \def\Pd      {\ensuremath{\mathrm{d}}\xspace}
 \def\Pe      {\ensuremath{\mathrm{e}}\xspace}
 \def\Pp      {\ensuremath{\mathrm{p}}\xspace}

 \def\Ps      {\ensuremath{\mathrm{s}}\xspace}
 
 \def\Pu      {\ensuremath{\mathrm{u}}\xspace}
 \def\thebaroffset{0.0em}
}
{
 
 \def\Pgamma      {\ensuremath{\gamma}\xspace}

 \def\Peta        {\ensuremath{\eta}\xspace}

 \def\Pmu         {\ensuremath{\mu}\xspace}
 \def\Pnu         {\ensuremath{\nu}\xspace}
 
 \def\Ppi         {\ensuremath{\pi}\xspace}
 
 \def\Prho        {\ensuremath{\rho}\xspace}

 \def\Pphi        {\ensuremath{\phi}\xspace}

 \def\Ppsi        {\ensuremath{\psi}\xspace}
 
 \mathchardef\PDelta="7101
 \mathchardef\PXi="7104
 \mathchardef\PLambda="7103
 \mathchardef\PSigma="7106
 \mathchardef\POmega="710A
 \mathchardef\PUpsilon="7107
 \mathchardef\PPi="7105
 \def\PB      {\ensuremath{B}\xspace}
 \def\PD      {\ensuremath{D}\xspace}
 \def\PJ      {\ensuremath{J}\xspace}
 \def\PK      {\ensuremath{K}\xspace}
 \def\Pb      {\ensuremath{b}\xspace}
 \def\Pc      {\ensuremath{c}\xspace}
 \def\Pd      {\ensuremath{d}\xspace}
 \def\Pe      {\ensuremath{e}\xspace}
 \def\Pp      {\ensuremath{p}\xspace}

 \def\Ps      {\ensuremath{s}\xspace}
 
 \def\Pu      {\ensuremath{u}\xspace}
 \def\thebaroffset{0.18em}
}
\newcommand{\offsetoverline}[2][\thebaroffset]{\kern #1\overline{\kern -#1 #2}}%

\makeatletter
\ifcase \@ptsize \relax% 10pt
  \newcommand{\miniscule}{\@setfontsize\miniscule{4}{5}}% \tiny: 5/6
\or% 11pt
  \newcommand{\miniscule}{\@setfontsize\miniscule{5}{6}}% \tiny: 6/7
\or% 12pt
  \newcommand{\miniscule}{\@setfontsize\miniscule{5}{6}}% \tiny: 6/7
\fi
\makeatother

\DeclareRobustCommand{\optbar}[1]{\shortstack{{\miniscule (\rule[.5ex]{1.25em}{.18mm})}
  \\ [-.7ex] $#1$}}

\def\ep         {{\ensuremath{\Pe^+}}\xspace}

\def\mup        {{\ensuremath{\Pmu^+}}\xspace}
\def\mun        {{\ensuremath{\Pmu^-}}\xspace} % muon negative (\mum is taken)

\def\neu        {{\ensuremath{\Pnu}}\xspace}

\def\neue       {{\ensuremath{\neu_e}}\xspace}

\def\g      {{\ensuremath{\Pgamma}}\xspace}

\def\uquark    {{\ensuremath{\Pu}}\xspace}
\def\uquarkbar {{\ensuremath{\overline \uquark}}\xspace}
\def\uubar     {{\ensuremath{\uquark\uquarkbar}}\xspace}
\def\dquark    {{\ensuremath{\Pd}}\xspace}
\def\dquarkbar {{\ensuremath{\overline \dquark}}\xspace}
\def\ddbar     {{\ensuremath{\dquark\dquarkbar}}\xspace}
\def\squark    {{\ensuremath{\Ps}}\xspace}
\def\squarkbar {{\ensuremath{\overline \squark}}\xspace}
\def\ssbar     {{\ensuremath{\squark\squarkbar}}\xspace}
\def\cquark    {{\ensuremath{\Pc}}\xspace}
\def\cquarkbar {{\ensuremath{\overline \cquark}}\xspace}
\def\ccbar     {{\ensuremath{\cquark\cquarkbar}}\xspace}
\def\bquark    {{\ensuremath{\Pb}}\xspace}
\def\bquarkbar {{\ensuremath{\overline \bquark}}\xspace}
\def\bbbar     {{\ensuremath{\bquark\bquarkbar}}\xspace}

\def\pion   {{\ensuremath{\Ppi}}\xspace}
\def\piz    {{\ensuremath{\pion^0}}\xspace}
\def\pip    {{\ensuremath{\pion^+}}\xspace}
\def\pim    {{\ensuremath{\pion^-}}\xspace}

\def\rhomeson {{\ensuremath{\Prho}}\xspace}
\def\rhoz     {{\ensuremath{\rhomeson^0}}\xspace}
\def\rhop     {{\ensuremath{\rhomeson^+}}\xspace}
\def\rhom     {{\ensuremath{\rhomeson^-}}\xspace}

\def\kaon    {{\ensuremath{\PK}}\xspace}
\def\KorKbar {\kern \thebaroffset\optbar{\kern -\thebaroffset \PK}{}\xspace}

\def\Kp      {{\ensuremath{\kaon^+}}\xspace}
\def\Km      {{\ensuremath{\kaon^-}}\xspace}

\def\Kstarz  {{\ensuremath{\kaon^{*0}}}\xspace}

\def\Kstar   {{\ensuremath{\kaon^*}}\xspace}

\def\Kstarp  {{\ensuremath{\kaon^{*+}}}\xspace}

\newcommand{\etaz}{\ensuremath{\Peta}\xspace}
\newcommand{\etapr}{\ensuremath{\Peta^{\prime}}\xspace}
\newcommand{\phiz}{\ensuremath{\Pphi}\xspace}

\def\D       {{\ensuremath{\PD}}\xspace}

\def\DorDbar {\kern \thebaroffset\optbar{\kern -\thebaroffset \PD}\xspace}

\def\Dp      {{\ensuremath{\D^+}}\xspace}
\def\Dm      {{\ensuremath{\D^-}}\xspace}

\def\DpDm    {\ensuremath{\Dp {\kern -0.16em \Dm}}\xspace}

\def\Dsp     {{\ensuremath{\D^+_\squark}}\xspace}

\def\B       {{\ensuremath{\PB}}\xspace}

\def\BorBbar {\kern \thebaroffset\optbar{\kern -\thebaroffset \PB}\xspace}
\def\Bz      {{\ensuremath{\B^0}}\xspace}

\def\Bd      {{\ensuremath{\B^0}}\xspace}

\def\BdorBdbar {\kern \thebaroffset\optbar{\kern -\thebaroffset \Bd}\xspace}
\def\Bu      {{\ensuremath{\B^+}}\xspace}

\def\Bs      {{\ensuremath{\B^0_\squark}}\xspace}

\def\BsorBsbar {\kern \thebaroffset\optbar{\kern -\thebaroffset \Bs}\xspace}

\def\Bds     {{\ensuremath{\B_{(\squark)}^0}}\xspace}

\def\jpsi     {{\ensuremath{{\PJ\mskip -3mu/\mskip -2mu\Ppsi}}}\xspace}
\def\psitwos  {{\ensuremath{\Ppsi{(2S)}}}\xspace}

\def\Y#1S{\ensuremath{\PUpsilon{(#1S)}}\xspace}

\def\proton      {{\ensuremath{\Pp}}\xspace}
\def\antiproton  {{\ensuremath{\overline \proton}}\xspace}

\def\LorLbar     {\kern \thebaroffset\optbar{\kern -\thebaroffset \PLambda}\xspace}

\def\BF         {{\ensuremath{\mathcal{B}}}\xspace}

\def\to                 {\ensuremath{\rightarrow}\xspace}

\def\CP                {{\ensuremath{C\!P}}\xspace}

\def\AT#1     {\ensuremath{A_{\mathrm{T}}^{#1}}\xspace}           % 2

\def\C#1      {\ensuremath{\mathcal{C}_{#1}}\xspace}                       % 9
\def\Cp#1     {\ensuremath{\mathcal{C}_{#1}^{'}}\xspace}                    % 7
\def\Ceff#1   {\ensuremath{\mathcal{C}_{#1}^{\mathrm{(eff)}}}\xspace}        % 9  
\def\Cpeff#1  {\ensuremath{\mathcal{C}_{#1}^{'\mathrm{(eff)}}}\xspace}       % 7
\def\Ope#1    {\ensuremath{\mathcal{O}_{#1}}\xspace}                       % 2
\def\Opep#1   {\ensuremath{\mathcal{O}_{#1}^{'}}\xspace}                    % 7

\newcommand{\ket}[1]{\ensuremath{|#1\rangle}}              % {b}
\newcommand{\nospaceunit}[1]{\ensuremath{\text{#1}}}
\newcommand{\aunit}[1]{\ensuremath{\text{\,#1}}}
\newcommand{\tev}{\aunit{Te\kern -0.1em V}\xspace}
\newcommand{\gev}{\aunit{Ge\kern -0.1em V}\xspace}
\newcommand{\mev}{\aunit{Me\kern -0.1em V}\xspace}
\newcommand{\kev}{\aunit{ke\kern -0.1em V}\xspace}
\newcommand{\ev}{\aunit{e\kern -0.1em V}\xspace}

\newcommand{\mevc}{\ensuremath{\aunit{Me\kern -0.1em V\!/}c}\xspace}
\newcommand{\gevc}{\ensuremath{\aunit{Ge\kern -0.1em V\!/}c}\xspace}
\newcommand{\mevcc}{\ensuremath{\aunit{Me\kern -0.1em V\!/}c^2}\xspace}
\newcommand{\gevcc}{\ensuremath{\aunit{Ge\kern -0.1em V\!/}c^2}\xspace}
\def\mum  {\ensuremath{\,\upmu\nospaceunit{m}}\xspace}

\def\fb   {\ensuremath{\aunit{fb}}\xspace}
\def\invfb   {\ensuremath{\fb^{-1}}\xspace}

\def\sec  {\ensuremath{\aunit{s}}\xspace}

\def\ps   {\ensuremath{\aunit{ps}}\xspace}

\def\gsim{{~\raise.15em\hbox{$>$}\kern-.85em
          \lower.35em\hbox{$\sim$}~}\xspace}
\def\lsim{{~\raise.15em\hbox{$<$}\kern-.85em
          \lower.35em\hbox{$\sim$}~}\xspace}

\def\sqs   {\ensuremath{\protect\sqrt{s}}\xspace}

\def\pt         {\ensuremath{p_{\mathrm{T}}}\xspace}

\def\ptot       {\ensuremath{p}\xspace}

\def\evtgen     {\mbox{\textsc{EvtGen}}\xspace}

\def\geant      {\mbox{\textsc{Geant4}}\xspace}

\def\photos     {\mbox{\textsc{Photos}}\xspace}

\def\pythia     {\mbox{\textsc{Pythia}}\xspace}

\def\tell1  {TELL1\xspace}
\def\ukl1   {UKL1\xspace}

\newcommand{\eg}{\mbox{\itshape e.g.}\xspace}

\newcommand{\lhcborcid}[1]{\href{https://orcid.org/#1}{\hspace*{0.1em}\raisebox{-0.45ex}{\includegraphics[width=1em]{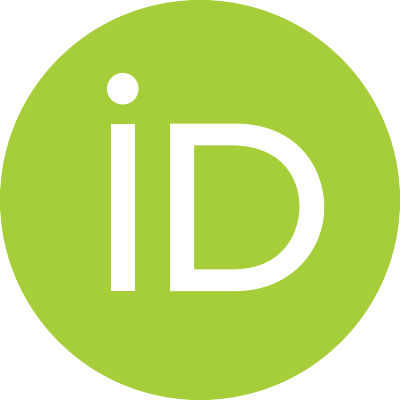}}}}

\hypersetup{
  colorlinks   = true, %Colours links instead of ugly boxes
  urlcolor     = blue, %Colour for external hyperlinks
  linkcolor    = blue, %Colour of internal links
  citecolor    = red   %Colour of citations
}

\ifEnableSectionTOCLinks
    \usepackage[explicit]{titlesec} % to change headings
    
    \let\oldcontentsline\contentsline
    \renewcommand

    \titleformat{\section}{\normalfont\Large\bf}{\hyperlink{tocsection.\thesection}{{\thesection} \parbox[t]{\dimexpr\textwidth-1pc}{#1}}}{1pc}{}

    \titleformat{\subsection}{\normalfont\bf}{\hyperlink{tocsubsection.\thesubsection}{{\thesubsection} \parbox[t]{\dimexpr\textwidth-1pc}{#1}}}{1pc}{}

    \titleformat{name=\section,numberless}[display]{}{}{0pt}{\normalfont\Huge\bfseries #1}
\fi

\usepackage{cite} % Allows for ranges in citations
\usepackage{mciteplus}
\usepackage{longtable} % only for template; not usually to be used in PAPERs
\usepackage{multirow}

\begin{document}

\newcommand{\phip}{{\ensuremath{\phi_{\rm{P}}}}\xspace}
\newcommand{\phig}{{\ensuremath{\phi_{\rm{G}}}}\xspace}
\newcommand{\phigeta}{{\ensuremath{\phi_{\rm{G}}^{\etaz}}}\xspace}
\newcommand{\rz}{{\ensuremath{R_{0}}}\xspace}
\newcommand{\rzp}{{\ensuremath{R_{0}^{'}}}\xspace}
\newcommand{\rd}{{\ensuremath{R_d}}\xspace}
\newcommand{\rs}{{\ensuremath{R_s}}\xspace}
\newcommand{\rds}{{\ensuremath{R_{d,s}}}\xspace}
\newcommand{\rdsgg}{{\ensuremath{R_{d,s}^{\g\g}}}\xspace}
\newcommand{\rdsg}{{\ensuremath{R_{d,s}^{\g}}}\xspace}
\newcommand{\rdg}{{\ensuremath{R_d}^{\g}}\xspace}
\newcommand{\rsg}{{\ensuremath{R_s}^{\g}}\xspace}
\newcommand{\rdgg}{{\ensuremath{R_d}^{\g\g}}\xspace}
\newcommand{\rsgg}{{\ensuremath{R_s}^{\g\g}}\xspace}
\newcommand{\etazpr}{\ensuremath{\Peta^{(\prime)}}\xspace}

%%%%%%%%%%%%%%%%%%%%%%%%%
%%%%% Title     %%%%%%%%%
%%%%%%%%%%%%%%%%%%%%%%%%%
\renewcommand{\thefootnote}{\fnsymbol{footnote}}
\setcounter{footnote}{1}

% %%%%%%% CHOOSE TITLE PAGE--------
%\onecolumn
\begin{titlepage}
\pagenumbering{roman}

% Header ---------------------------------------------------
\vspace*{-1.5cm}
\centerline{\large EUROPEAN ORGANIZATION FOR NUCLEAR RESEARCH (CERN)}
\vspace*{1.5cm}
\noindent
\begin{tabular*}{\linewidth}{lc@{\extracolsep{\fill}}r@{\extracolsep{0pt}}}
\ifthenelse{\boolean{pdflatex}}% Logo format choice
{\vspace*{-1.5cm}\mbox{\!\!\!\includegraphics[width=.14\textwidth]{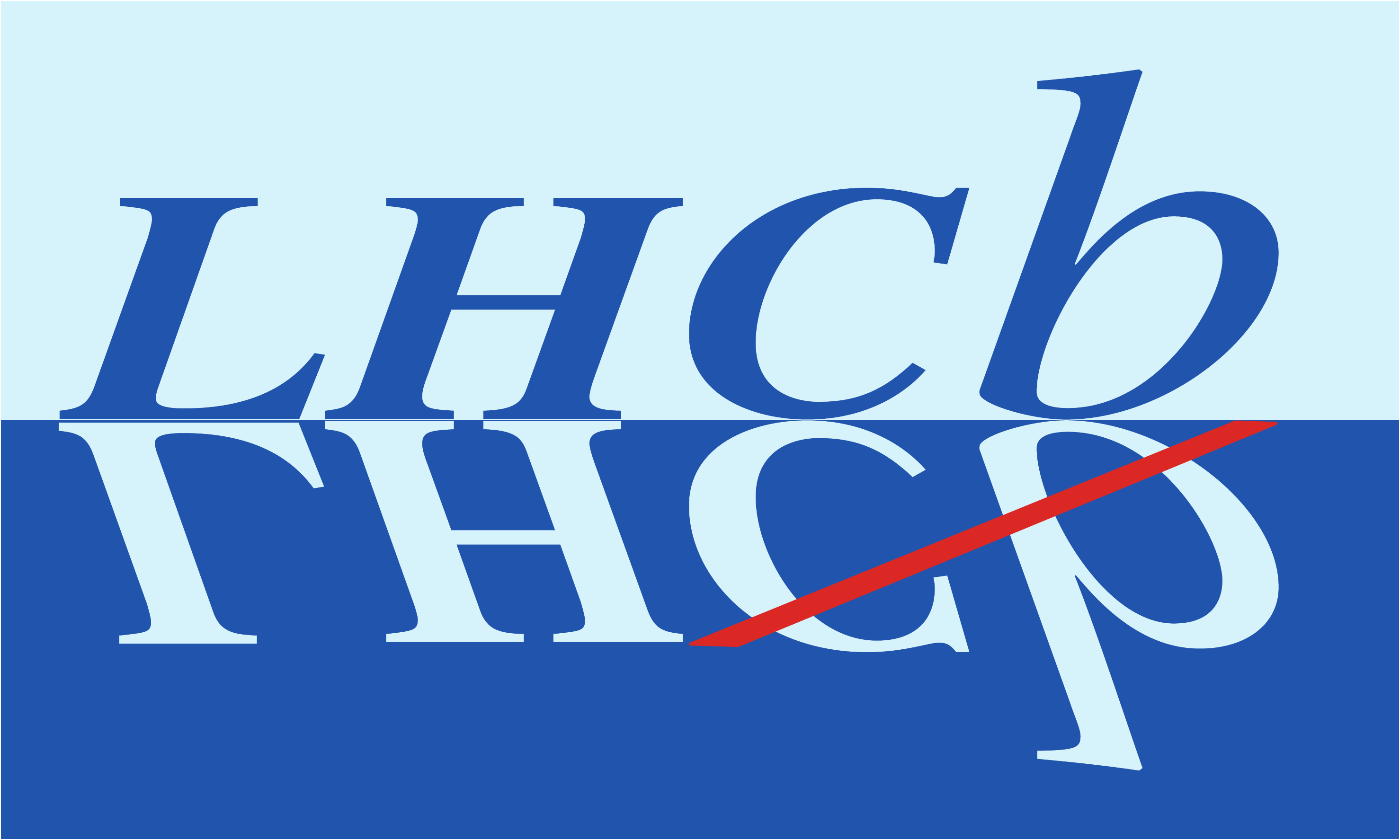}} & &}%
{\vspace*{-1.2cm}\mbox{\!\!\!\includegraphics[width=.12\textwidth]{lhcb-logo.pdf}} & &}%
\\
 & & CERN-EP-2025-148 \\  % ID 
 & & LHCb-PAPER-2025-025 \\  % ID 
 & & 3 November 2025 \\ % Date
 & & \\
\end{tabular*}

\vspace*{0.75cm}

{\normalfont\bfseries\boldmath\huge
\begin{center}
% DO NOT EDIT HERE. Instead edit macro in main.tex to keep metadata correct
  \papertitle 
\end{center}
}

\vspace*{0.75cm}

\begin{center}
\paperauthors\footnote{Authors are listed at the end of this paper.}
\end{center}

\vspace{\fill}

\begin{abstract}
  \noindent
  Branching fraction ratios between the decays $\Bds\to\jpsi\etazpr$ are measured using proton-proton collision data collected by the LHCb experiment at centre-of-mass energies of 7, 8 and 13\tev, corresponding to an integrated luminosity of 9\invfb.
  The measured ratios of these branching fractions are
    \begin{displaymath}
        \begin{split}
        \frac{\mathcal{B}(\Bz\to\jpsi \etapr)}{\mathcal{B}(\Bz\to\jpsi \etaz)}		&= 0.48 \pm 0.06 \pm 0.02 \pm 0.01, \\	
        \frac{\mathcal{B}(\Bs\to\jpsi \etapr)}{\mathcal{B}(\Bs\to\jpsi \etaz)}	&= 0.80 \pm 0.02 \pm 0.02 \pm 0.01, \\	
        \end{split}
    \end{displaymath}
  where the uncertainties are statistical, systematic and related to the precision of the \etazpr branching fractions, respectively. They are used to constrain the \etaz/\etapr mixing angle, \phip, and to probe the presence of a possible glueball component in the \etapr meson, described by the gluonic mixing angle \phig. The obtained results are
    \begin{displaymath}
        \begin{split}
        \phip &= (41.6^{+1.0}_{-1.2})^\circ,\\
        \phig &= (28.1^{+3.9}_{-4.0})^\circ,\\
        \end{split}
    \end{displaymath}
  where the uncertainties are statistically dominated.
  While the value of \phip is compatible with existing experimental determinations and theoretical calculations, the angle \phig differs from zero by more than four standard deviations, which points to a substantial glueball component in the \etapr meson and/or unexpectedly large contributions from gluon-mediated processes in these decays.
  The absolute branching fractions are also measured relative to that of the well-established $\Bs\to\jpsi\phiz$ decay, which serves as the normalisation channel. These results supersede the previous LHCb measurements and are the most precise to date.
\end{abstract}

\vspace*{0.5cm}

\begin{center}
  Published in JHEP 10 (2025) 113
\end{center}

{\footnotesize 
\centerline{\copyright~\papercopyright. \href{\paperlicenceurl}{\paperlicence}.}}
\vspace*{2mm}

\end{titlepage}

\newpage
\setcounter{page}{2}
\mbox{~}

%\twocolumn
% %%%%%%%%%%%%% ---------

\renewcommand{\thefootnote}{\arabic{footnote}}
\setcounter{footnote}{0}

%%%%%%%%%%%%%%%%%%%%%%%%%%%%%%%%
%%%%%  Table of Content   %%%%%%
%%%%%%%%%%%%%%%%%%%%%%%%%%%%%%%%
%%%% Uncomment if desired
%\tableofcontents

\cleardoublepage

%%%%%%%%%%%%%%%%%%%%%%%%%
%%%%% Main text %%%%%%%%%
%%%%%%%%%%%%%%%%%%%%%%%%%

\pagestyle{plain} % restore page numbers for the main text
\setcounter{page}{1}
\pagenumbering{arabic}

%% Uncomment during review phase. 
%% Comment before a final submission.
%% \linenumbers

%% This is the main body
%% It is useful to have a single file so comments are not missed in overleaf.
\section{Introduction}
\label{sec:intro}

Decays of the \etazpr mesons provide a unique, flavour-conserving laboratory to test low-energy quantum chromodynamics (QCD) and search for physics beyond the Standard Model~\cite{Kubis22}.
Their properties reflect the low-energy dynamics of quarks and gluons --- the fundamental degrees of freedom of QCD --- and remain challenging to predict due to the effects of confinement.
The constituent quark model~\cite{Amsler:2018zkm} describes the \etaz and \etapr pseudoscalar mesons as linear combinations of the isospin-0 octet and singlet SU(3)-flavour states, where mixing is a consequence of the larger mass of the $s$ quark relative to the $u$ and $d$ quarks.
These combinations are parametrised by a mixing angle, whose predicted value lies between $-25^{\circ}$ and $-10^{\circ}$, defining the \etapr (\etaz) meson as mostly singlet (octet) state.

More elaborate approaches to calculate the properties of light hadrons rely on the symmetry properties of QCD, such as the approximate chiral symmetry among light quarks~\cite{Scherer12}.
This spontaneously broken symmetry generates the octet of pseudoscalar mesons and forms the basis for chiral perturbation theory ($\chi\textrm{PT}$)~ \cite{Feldmann99}.
In the large-number-of-colours limit of QCD, $\chi\textrm{PT}$ can be extended to include the singlet state, providing a consistent framework to describe the mixing parameters, and showing that a one-angle description is only valid at leading order in perturbation theory~\cite{Guo15,Bickert17}.

Phenomenological analyses of different physics processes, such as transitions between light and heavy hadrons (\eg $\Dsp\to\etazpr\ep\neue$,\footnote{Inclusion of charge-conjugate processes is implied
throughout.} $V\to\etazpr\g$ with $V=\jpsi,\phiz$), diphoton decays (\eg $\etazpr\to\g\g$) and annihilation processes (\eg $p\antiproton\to PM$ with $P$ a pseudoscalar meson and $M=\piz,\etaz,\omega$)~\cite{Feldmann98,Kou01,Thomas07,Escribano99,Escribano05}, advocate the use of two angles to describe \etaz/\etapr mixing in the octet/singlet basis.
These analyses also emphasise the special role of the quark basis, spanned by the states $\ket{\eta_q}=1/\sqrt{2}(\ket{\uubar}+\ket{\ddbar})$ and $\ket{\eta_s}=\ket{\ssbar}$, where these two angles take compatible values and are replaced by a single angle, \phip, with a value of around 40$^{\circ}$ (see Ref.~\cite{LHCb-PAPER-2014-056} for a review of recent determinations).
First-principle calculations on the lattice are becoming competitive in precision and agree with these determinations~\cite{Bali21,Kordov21,ETM18}.
Possible \etazpr mixing with \piz mesons or charmonium has also been investigated and found to be negligible~\cite{Feldmann98,Escribano20}.

Another intriguing aspect of \etazpr mesons is their possible gluonic content and the connection with glueballs, the colour singlet states made of two or more gluons predicted by QCD~\cite{MATHIEU_2009}.
Interestingly, the QCD anomaly~\cite{Adler69,BellJackiw69,Greiner07} contributes to the singlet mass by a purely gluonic component which, due to mixing, mostly affects the \etapr meson ~\cite{ATWOOD1997150}.
Because of its gluonic content, the \etapr meson may mix with glueballs~\cite{BALL1996367,Zhang_2018}, hence changing the rates of physics processes where it is involved (\eg $\Bu\to\Kp\etapr$~\cite{Donato12}, $\Bds\to\jpsi\etapr$~\cite{Fleischer11}), and providing indirect constraints on the mass of the lightest pseudoscalar glueball~\cite{PhysRevD.79.014024}.
Current experimental results suggest a negligible glueball component in the $\eta'$ meson, although earlier reports indicated potentially large contributions with limited statistical significance~\cite{Escribano07,Kou01,Escribano09,Liu12,Donato12,Ambrosino09}.
Potential glueball components to the \etaz meson have also been investigated using \jpsi decays~\cite{Escribano09} and were not found to be significant.

At the quark level, $\Bds\to\jpsi\etazpr$ decays proceed dominantly through the tree-level transition $b\to\ccbar q$, where $q$ is either a $d$ or an $s$ quark that forms an \etazpr meson with the spectator quark.
Neglecting decay topologies where the spectator quark participates in the flavour-changing interaction and the \etazpr or \jpsi meson is produced by gluons, simple relations between decay amplitudes can be derived which relate the relative decay rates to the mixing angle \phip~\cite{Datta02}.
These relations allow to test the validity of the one-angle \etaz/\etapr mixing scheme (standard mixing) or the presence of the aforementioned gluon-mediated processes coupling differently to the two mesons.
As proposed in Ref.~\cite{Fleischer11}, the standard mixing scheme can be extended to incorporate a glueball component to the \etapr wave function.
In the quark basis, the physical states are written as
\begin{equation}
\begin{split}
\ket{\eta}\phantom{'} &= \cos\phip\, \ket{\eta_q} - \sin\phip\, \ket{\eta_s},\\
\ket{\etapr} &= \cos\phig ( \sin\phip\, \ket{\eta_q} + \cos\phip\, \ket{\eta_s} ) + \sin\phig\,\ket{gg},\\
\end{split}
\label{eq:eta_wave}
\end{equation}
where $\ket{gg}$ denotes a glueball state and \phig is the gluonic angle.
Projecting out the physical states to the basis states, the following relations between branching fractions are obtained:
\begin{equation}
  \rd \equiv \frac{\BF (\Bz\to\jpsi\etapr)}{\BF (\Bz\to\jpsi\etaz)} \cdot \frac{\Phi^{3}(\Bz\to\jpsi\etaz)}{\Phi^{3}(\Bz\to\jpsi\etapr)} = \tan^{2} \phip \cdot \cos^{2} \phig,
\label{eq:intro_rd}
\end{equation}
\begin{equation}
  \rs \equiv \frac{\BF (\Bs\to\jpsi\etapr)}{\BF (\Bs\to\jpsi\etaz)} \cdot \frac{\Phi^{3}(\Bs\to\jpsi\etaz)}{\Phi^{3}(\Bs\to\jpsi\etapr)} = \cot^{2} \phip \cdot \cos^{2} \phig,
\label{eq:intro_rs}
\end{equation}
where the $\Phi$ terms are phase-space factors defined as
\begin{equation}
  \Phi(\Bds\to\jpsi\etazpr) = \sqrt{ \Biggr[ 1 - \biggl( \frac{m_{\jpsi} - m_{\etazpr}}{m_{\Bds}} \biggl)^{2} \Biggr] \Biggr[ 1 - \biggl( \frac{m_{\jpsi} + m_{\etazpr}}{m_{\Bds}} \biggl)^{2} \Biggr] },
\label{eq:intro_phsp}
\end{equation}
that are raised to the power three in Eqs.\,\ref{eq:intro_rd} and \ref{eq:intro_rs} to account for the spin configuration of the decay.
The values for the angles can then be obtained from a combination of the \rds ratios, as
\begin{equation}
  \tan^{4}\phip = \rd \,/ \rs,
\label{eq:intro_phip}
\end{equation}
\begin{equation}
  \cos^{4}\phig = \rd \cdot \rs,
\label{eq:intro_phig}
\end{equation}
where, in the case of standard mixing, $\phig=0$ and $\rd \cdot \rs = 1$~\cite{Datta02}.
Similarly, one can calculate the following ratios which only depend on \phip,
\begin{equation}
  R_{\etaz} \equiv \frac{\BF (\Bz\to\jpsi\etaz)}{\BF (\Bs\to\jpsi\etaz)} \cdot \frac{\Phi^{3}(\Bs\to\jpsi\etaz)}{\Phi^{3}(\Bz\to\jpsi\etaz)} = \frac{\tau_{\Bz} m_{\Bz}}{\tau_{\Bs} m_{\Bs}} \bigg|\frac{V_{cd}}{V_{cs}}\bigg|^{2} \frac{\cot^{2}{\phip}}{2},
\label{eq:intro_retaz}
\end{equation}
\begin{equation}
  R_{\etapr} \equiv \frac{\BF (\Bz\to\jpsi\etapr)}{\BF (\Bs\to\jpsi\etapr)} \cdot \frac{\Phi^{3}(\Bs\to\jpsi\etapr)}{\Phi^{3}(\Bz\to\jpsi\etapr)} = \frac{\tau_{\Bz} m_{\Bz}}{\tau_{\Bs} m_{\Bs}} \bigg|\frac{V_{cd}}{V_{cs}}\bigg|^{2} \frac{\tan^{2}{\phip}}{2},
\label{eq:intro_retapr}
\end{equation}
which hold up to SU(3)-breaking corrections, and where $m$, $\tau$ and $V_{ij}$ are the \Bds mass and lifetime, and CKM matrix elements, respectively~\cite{Datta02}.

Initial experimental results were reported in 2012 by the Belle collaboration for the decay $\Bz\to\jpsi\etaz$ and the two modes $\Bs\to\jpsi\etaz$ and $\Bs\to\jpsi\etapr$~\cite{PhysRevLett.108.181808,PhysRevD.85.091102}.
These measurements were interpreted as evidence for a sizeable glueball component in the \etapr meson, based on comparison with theoretical predictions for the branching fractions~\cite{Liu12}.
Results for the two \Bs modes were first reported by the LHCb collaboration using proton-proton ($pp$) collision data corresponding to an integrated luminosity of $1\invfb$~\cite{LHCb-PAPER-2012-022}.
A second LHCb analysis, based on a larger integrated luminosity of $3\invfb$, measured the four decay modes and found no evidence for a glueball component. The study determined the mixing angles to be $\phip = (43.5^{+1.4}_{-2.8})^{\circ}$ and $\phig = (0.0 \pm 24.6)^{\circ}$~\cite{LHCb-PAPER-2014-056}.

This analysis uses the $pp$ sample recorded by LHCb at centre-of-mass energies of 7, 8 and 13\,\tev, corresponding to an integrated luminosity of 9\,\invfb, to determine the branching fraction ratios between $\Bds\to\jpsi\etapr$ and $\Bds\to\jpsi\etaz$ decays, and subsequently the mixing angles. Ratios of branching fractions between the two $B$ flavours for a given final state are also determined (Eqs.\,\ref{eq:intro_retaz} and \ref{eq:intro_retapr}), although they are sensitive to \phip only.
The observables are measured as
\begin{equation}
  \rds = \frac{N(\Bds\to\jpsi\etapr)}{N(\Bds\to\jpsi\etaz)} \cdot \frac{\varepsilon(\Bds\to\jpsi\etaz)}{\varepsilon(\Bds\to\jpsi\etapr)} \cdot \frac{\BF (\etaz\to f_{\etaz})}{\BF (\etapr\to f_{\etapr})} \cdot \frac{\Phi^{3}(\Bds\to\jpsi\etaz)}{\Phi^{3}(\Bds\to\jpsi\etapr)},
\label{eq:rds_exp}
\end{equation}
where the secondary branching fractions $\mathcal{B}$ for the \etazpr decays to the final states $f_{\etazpr}$ are used, and
\begin{equation}
  R_{\etazpr} = \frac{N(\Bz\to\jpsi\etazpr)}{N(\Bs\to\jpsi\etazpr)} \cdot \frac{\varepsilon(\Bs\to\jpsi\etazpr)}{\varepsilon(\Bz\to\jpsi\etazpr)} \cdot \frac{f_s}{f_d} \cdot \frac{\Phi^{3}(\Bs\to\jpsi\etazpr)}{\Phi^{3}(\Bz\to\jpsi\etazpr)},
\label{eq:retazpr_exp}
\end{equation}
where $f_s/f_d$ is the ratio of $b$-quark fragmentation fractions governing the relative production of \Bds mesons.

Absolute branching fractions are calculated using the decay $\Bs\to\jpsi\phiz$ as a normalisation channel.
As suggested in Ref.~\cite{Fleischer11}, a complementary determination of the angle \phip is possible using the measurement of $\mathcal{B}(\Bz\to\jpsi\etaz)$ and the ratio
\begin{equation}
  \rz \equiv \frac{\BF (\Bz\to\jpsi\etaz)}{\BF (\Bz\to\jpsi\piz)} \cdot \frac{\Phi^{3}(\Bz\to\jpsi\piz)}{\Phi^{3}(\Bz\to\jpsi\etaz)} = \cos^{2} \phip.
\label{eq:intro_r0}
\end{equation}
This determination uses the known branching fractions of $\Bz\to\jpsi\piz$ and $\Bs\to\jpsi\phiz$ decays (the latter being used to determine $\BF(\Bz\to\jpsi\etaz)$) and is therefore less precise than that based on the \rds ratios.
It is nevertheless interesting as the ratio \rz is sensitive to a possible glueball component in the \etaz meson which would modify Eq.\,\ref{eq:intro_r0} as $\rz = \cos^{2} \phip \cdot \cos^{2} \phigeta$, where $\phigeta$ is the corresponding mixing angle.

This analysis further improves upon the sensitivity of the previous LHCb analysis~\cite{LHCb-PAPER-2014-056} by incorporating a second final state for the reconstruction of the $\etazpr$ mesons, allowing for two independent determinations of the branching fraction ratios, which are subsequently combined.
The paper is organised as follows. A brief description of the experimental apparatus is given in Sec.~\ref{sec:detdesc} and the selection of signal and normalisation candidates are detailed in Sec.~\ref{sec:sel}. The selection efficiencies are calculated using simulation in Sec.~\ref{sec:eff}, while the yields of the various decays are determined from fits to the data, as explained in Sec.~\ref{sec:yields}. Systematic uncertainties affecting the measurements of the branching fraction ratios and of the angles are presented in Sec.~\ref{sec:syst}. Results for these observables are presented and briefly discussed in Sec.~\ref{sec:results}.

\section{Detector description}
\label{sec:detdesc}

The \lhcb detector~\cite{LHCb-DP-2012-002,LHCb-DP-2014-002} is a single-arm forward spectrometer covering the \mbox{pseudorapidity} range $2<\eta <5$, designed for the study of particles containing beauty or charm quarks. It includes a high-precision tracking system consisting of a silicon-strip vertex detector (VELO) surrounding the $pp$-interaction region, a large-area silicon-strip detector (TT) located upstream of a dipole magnet with a bending power of approximately $4{\mathrm{\,T\,m}}$, and three stations of silicon-strip detectors and straw drift tubes placed downstream of the magnet. The tracking system provides a measurement of the momentum, \ptot, of charged particles with a relative uncertainty that varies from 0.5\,\% at low momentum to 1.0\,\% at 200\gevc. 
The minimum distance of a track to a primary vertex (PV), the impact parameter (IP), is measured with a resolution of $(15+29/\pt)\mum$, where \pt is the component of the momentum transverse to the beam, in\,\gevc.
Various charged hadrons are distinguished using information from two ring-imaging Cherenkov detectors. In addition, photons, electrons, and hadrons are identified by a calorimeter system consisting of scintillating-pad and preshower detectors, an electromagnetic and a hadronic calorimeter.
The electromagnetic calorimeter response is calibrated using samples of $\piz \rightarrow \gamma \gamma$ decays~\cite{LHCb-DP-2020-001}.
Muons are identified by a system composed of alternating layers of iron and multiwire proportional chambers.

The online event selection is performed by a trigger, which consists of a hardware stage followed by a two-level software stage~\cite{LHCb-DP-2019-001}. An alignment and calibration of the detector is performed in near real-time with the results used in the software trigger~\cite{LHCb-PROC-2015-011}. The same alignment and calibration information is propagated to the offline reconstruction, ensuring consistent information between the trigger and offline software. In this analysis, candidate events are required to pass the hardware trigger, which selects muon and dimuon candidates with high transverse momenta using information from the muon system. The first stage of the software trigger performs a partial event reconstruction and requires events to have two well-identified oppositely charged muons with an invariant mass larger than $2700 \mevcc$. The second stage performs a  full event reconstruction.  Events are retained for further  processing if they contain a displaced $\jpsi \rightarrow \mu^+ \mu^-$
candidate. The decay vertex is required to be well separated from each reconstructed PV of the $pp$ interaction by requiring the distance between the PV and the $\jpsi$ decay vertex  divided by its uncertainty to be greater than three. 

Simulated $pp$ collisions are generated using \pythia~\cite{Sjostrand:2006za,*Sjostrand:2007gs} with a specific \lhcb configuration~\cite{LHCb-PROC-2010-056}.  Decays of hadronic particles are described by \evtgen~\cite{Lange:2001uf}, in which final-state radiation is generated using \photos~\cite{davidson2015photos}. The interaction of the generated particles with the detector, and its response, are implemented using the \geant toolkit~\cite{Allison:2006ve,*Agostinelli:2002hh} as described in Ref.~\cite{LHCb-PROC-2011-006}. 
The production of some samples is based on a computing-efficient model where the underlying $pp$ interaction is reused multiple times, with an independently generated signal decay for each~\cite{LHCb-DP-2018-004}.
The resulting selection efficiencies are found to be compatible with those based on the default production model, with slightly larger uncertainties resulting from the residual correlation between reused events.

\section{Selections}
\label{sec:sel}

Signal candidates are reconstructed using $\jpsi\to\mup\mun$ decays, and the $\pip\pim\g$ (one-photon) and $\pip\pim\g\g$ (two-photon) final states of the \etazpr decays, namely: $\etapr\to\rhoz(\to\pip\pim)\g$,\footnote{Throughout the paper, the symbol $\rho$ denotes the $\rho(770)$ resonance, and the presence of the \rhoz resonance in the \etapr decay is omitted.} $\etaz\to\pip\pim\g$, $\etapr\to\pip\pim\etaz(\to\g\g)$ and $\etaz\to\pip\pim\piz(\to\g\g)$.
Due to worse mass resolution and the small $\etapr\to\g\g$ branching fraction \cite{PDG2024}, $\Bds\to\jpsi\etazpr(\to\g\g)$ decays are not studied.
For the normalisation channel $\Bs\to\jpsi\phiz$, the \phiz meson is reconstructed using the decay to $\pip\pim\piz$, with $\piz\to\g\g$.

The two muons must satisfy $\pt>500\mevc$, pass a loose particle-identification (PID) requirement and be unlikely to originate from a PV. Furthermore, their combination should form a good vertex that is significantly displaced from any PV, have a mass within $\pm100\mevcc$ of the known \jpsi mass~\cite{PDG2024} and an IP value larger than 18\,\mum.
The \jpsi candidates are combined with pions and photons.
The pions are required to have $\pt>250\mevc$ and be unlikely to originate from a PV.
Similar to \jpsi candidates, a requirement is applied to the quality of the dipion vertex.
In the $\jpsi\etapr(\to\pip\pim\g)$ final state only, the dipion mass is required to be larger than 500\,\mevcc.

In the two-photon modes, photons must satisfy $\pt>200\mevc$ and be identified as well-isolated clusters in the electromagnetic calorimeter based on the response of a neural-net estimator trained to reject hadronic background.
The diphoton combination must have $\pt > 500\mevc$ ($1000\mevc$) for $\etaz\to\pip\pim\piz$ ($\etapr\to\pip\pim\etaz$) decays, and a mass within $\pm100\mevcc$ ($\pm155\mevcc$) of the known \piz (\etaz) mass~\cite{PDG2024}. These mass windows are significantly broader than the experimental resolution, as required for the evaluation of systematic uncertainties later in the analysis.
Diphoton and dipion candidates are then combined to form \etazpr candidates, which are required to have $\pt>2000\mevc$.
The mass of \etaz (\etapr) candidates must lie in the range \mbox{500--650\,\mevcc} (800--1200\,\mevcc), which is also significantly broader than the experimental resolution.

In the one-photon modes, the photon \pt requirement is increased to 500\,\mevc to reduce combinatorial background and the same requirement on cluster isolation is used. Moreover, a relatively wide mass window of 400--750\,\mevcc (800--1200\,\mevcc) is chosen for the \etaz (\etapr) candidates, which must also have $\pt>2000\mevc$.

The \etazpr and \jpsi candidates are then combined and the four-track vertex is required to have good fit $\chi^{2}$ and be significantly displaced from the PV that fits best to the flight direction of the $\Bds$ candidate.
Additionally, these candidates must have a mass in the range 4800--6200\,\mevcc and a \pt larger than 1800\,\mevc.
To improve the \Bds mass resolution, a kinematic fit~\cite{Hulsbergen:2005pu} is performed where each candidate is required to originate from the PV and the reconstructed masses of the intermediate narrow resonances (\jpsi, \etazpr, \phiz and \piz) are constrained to their known values.
For the decay modes with two photons, a second kinematic fit is performed where the mass constraint
on the $\pip\pim\g\g$ system is removed. This allows to improve the \etazpr mass resolution and hence to reduce combinatorial background below the corresponding mass peaks, as described later in this section.

Further selections are based on a boosted decision tree (BDT)~\cite{Breiman,AdaBoost} implemented in the TMVA toolkit~\cite{Hocker:2007ht}. This BDT classifier uses twelve variables which are accurately described by the simulation. These are related to the decay vertex, the kinematics of final-state particles, as well as the number of particles from the underlying event that are reconstructed close to the signal momentum direction, and their \pt.
A total of eight classifiers are trained, one for each of the four signal final states and run periods (Run~1 from 2011 to 2012 and Run~2 from 2015 to 2018).
The training samples consist of simulated \Bs decays and same-sign  data candidates ($\jpsi\pip\pip\g(\g)$) for signal and background, respectively.

For a given final state, the BDT threshold and the size of the $\pip\pim\g(\g)$ and diphoton mass windows are chosen to maximise the significance of the \Bz signal defined as $S/\sqrt{S+B}$. Here, $S$ and $B$ are the expected \Bz yield and background contamination in the \Bz signal region, respectively, corresponding to approximately three times the \Bz mass resolution.
They are determined using a fit to the $\jpsi\pip\pim\g(\g)$ mass distribution in data where the \Bz, \Bs and background yields are free to vary.
Given the large uncertainty in the \Bz yields, the value of $S$ is inferred from the \Bs yield, the Run~1 value of \phip, and other normalisation factors.
At the chosen working points, the BDT selections reject 95--99\% of background and retain 70--86\% of the signal.
The chosen mass windows correspond to $\pm2\sigma$ in the two-photon modes, where $\sigma$ represents the experimental resolution. In the one-photon modes, the \etaz (\etapr) mass window corresponds to $\pm1\sigma$ ($\pm1.5\sigma$).

Contamination from $\Bds\to\jpsi\pip\pim$ decays can occur when the final-state particles are combined with a reconstructed photon in the event; and similarly for the decays $\Bz\to\jpsi\Kp\pim$, where, in addition, the kaon is misidentified as a pion.
These backgrounds are only significant in the one-photon modes and are removed by rejecting signal candidates with a $\jpsi h^{+} \pim$ mass, where $h^{+}$ is a kaon or a pion, in the vicinity of the known \Bds mass. The effect of these vetoes on the shape of the combinatorial background mass distribution is investigated using same-sign data and is found to be negligible.

The selection of normalisation $\Bs\to\jpsi\phiz(\to\pip\pim\piz)$ decays is largely aligned with that of $\Bds\to\jpsi\etaz(\to\pip\pim\piz)$ decays.
Due to the larger phase space of the \phiz decay, the normalisation sample exhibits higher combinatorial background which is mitigated by a more stringent \piz \pt requirement of $1000\mevc$.
Furthermore, dedicated BDT classifiers are trained using simulation where the \phiz meson decays to $\rhop\pim$, $\rhom\pip$ or $\rhoz\piz$.
The BDT threshold and the \phiz and \piz mass requirements are similar to those in the signal mode to minimise systematic uncertainties.

After all selections, about 1--2\% of events contain multiple candidates, primarily due to the signal decays where one photon is replaced by another photon from the event. These candidates are not removed and are instead modelled in the fit, as detailed in Sec.~\ref{sec:yields_fit}.

\section{Estimation of efficiencies}
\label{sec:eff}

Selection efficiencies for the \Bs modes are calculated for each year of data taking using simulation, and then averaged according to the corresponding integrated luminosity and \bbbar production cross-section.
With a smaller set of simulated samples, the \Bz efficiency averages are calculated as the product of the \Bs averages and the $\Bz/\Bs$ efficiency ratios, where those ratios are obtained for the same data-taking conditions. The associated uncertainty is discussed in Sec.~\ref{sec:syst}.

\begin{figure}[tb]
  \centering
    \includegraphics[width=0.45\textwidth]{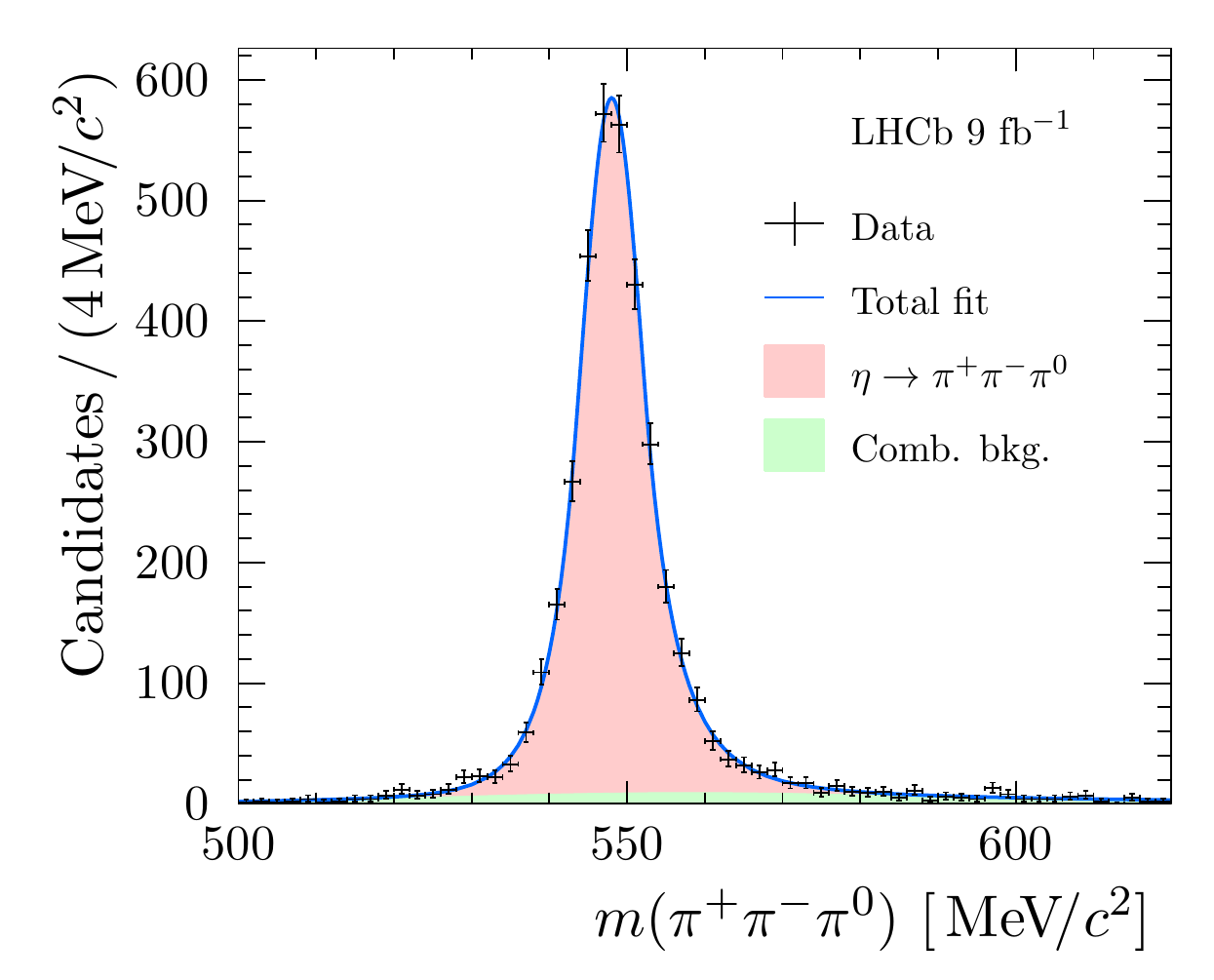}
    \includegraphics[width=0.45\textwidth]{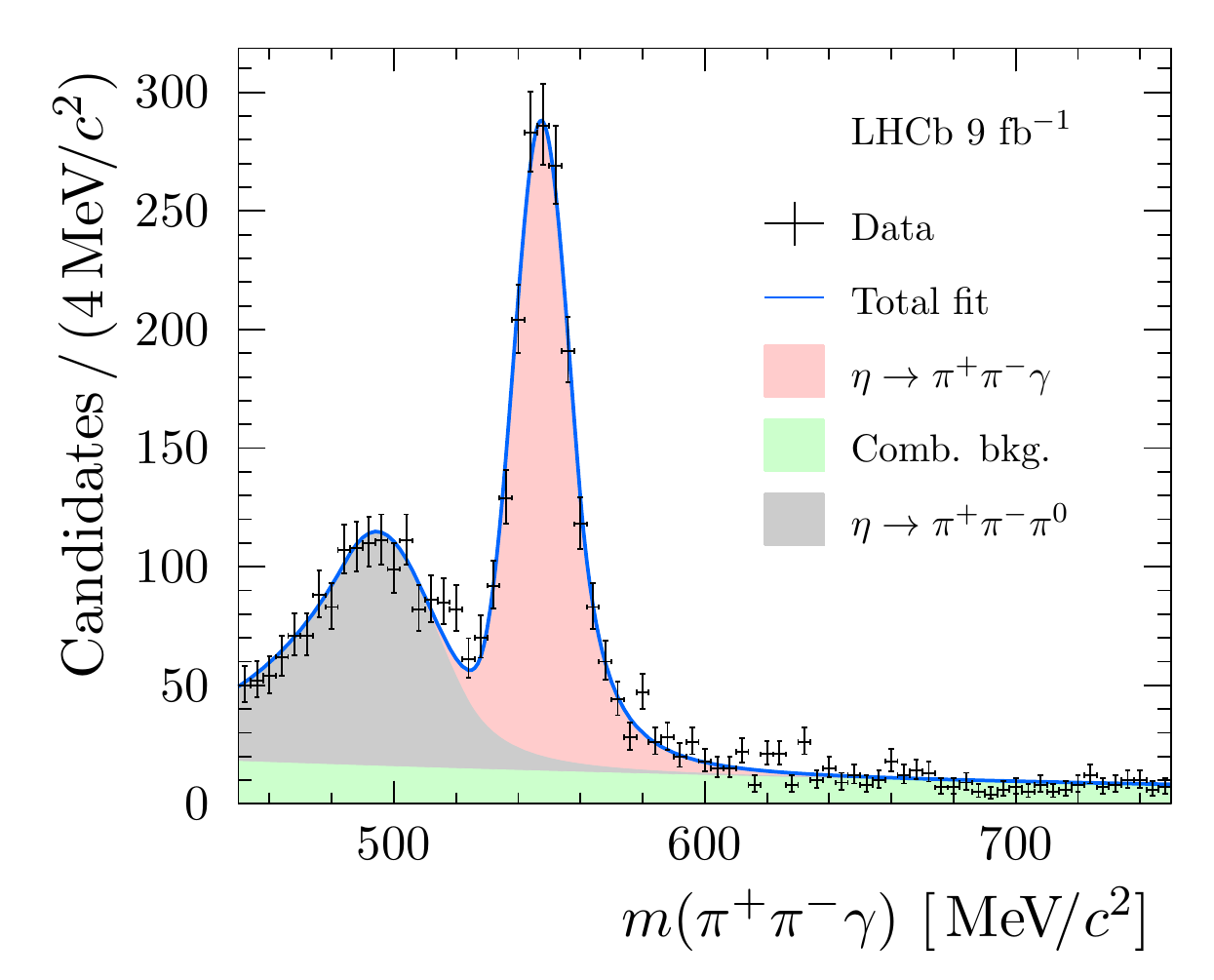}
    \includegraphics[width=0.45\textwidth]{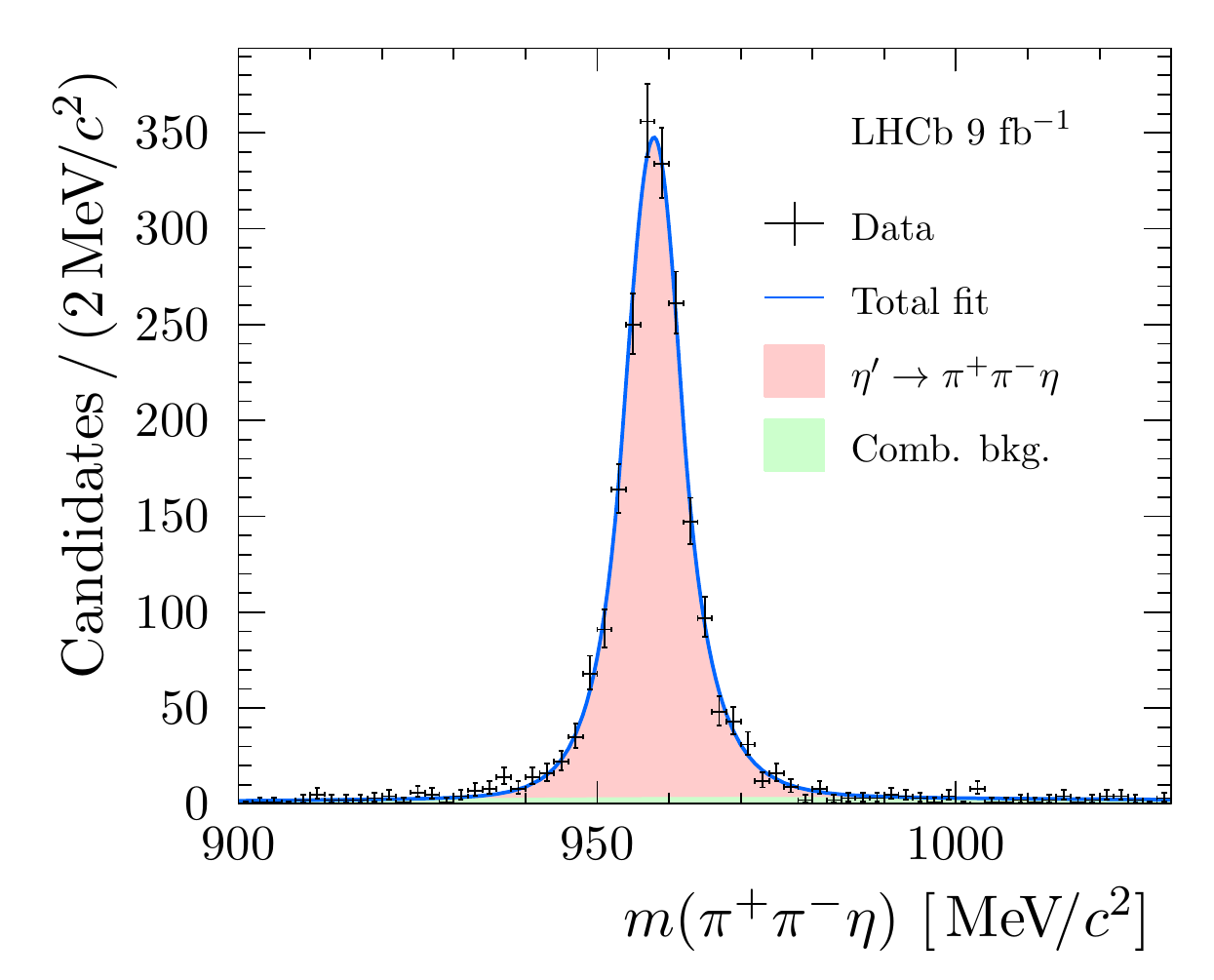}
    \includegraphics[width=0.45\textwidth]{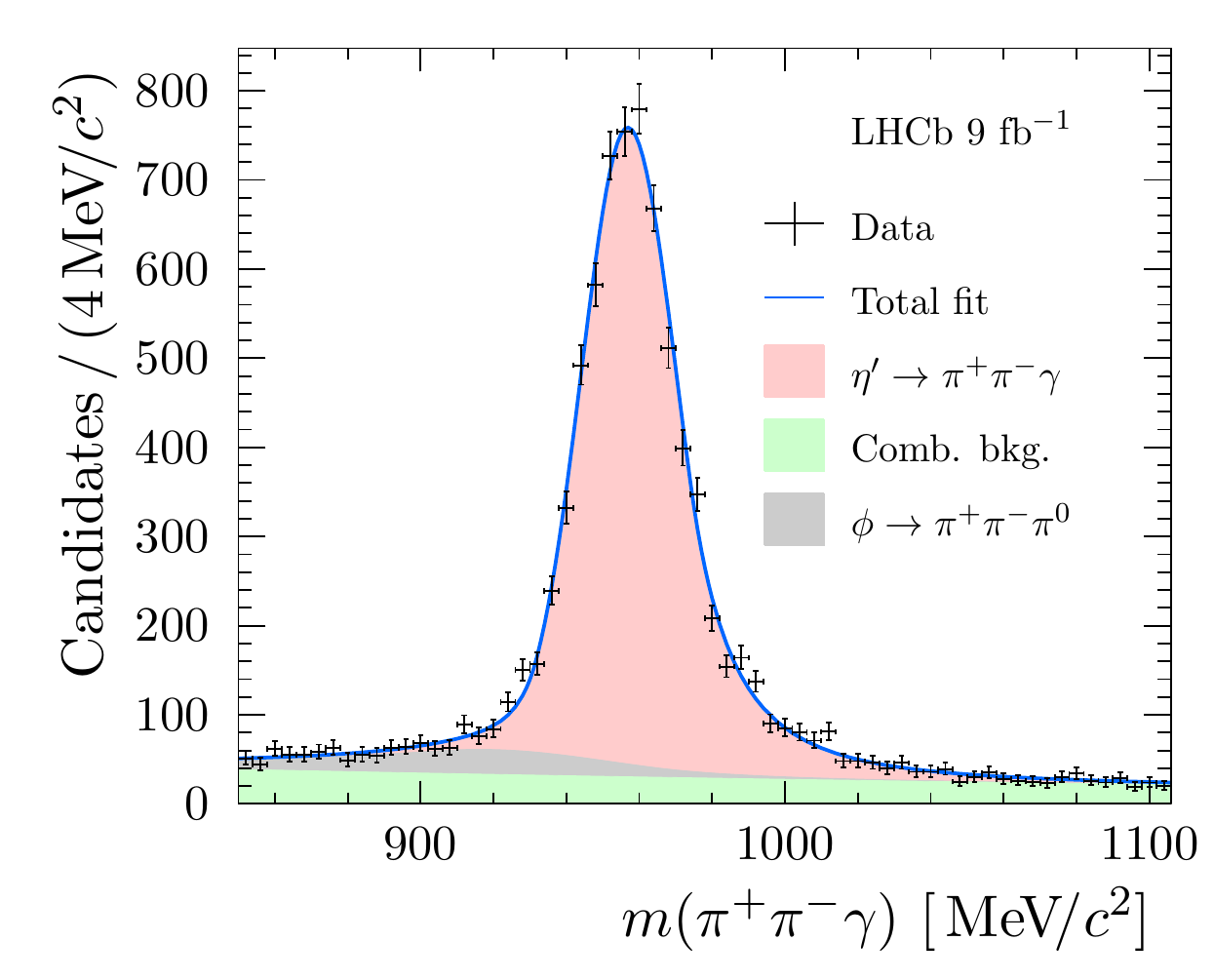}
    \includegraphics[width=0.45\textwidth]{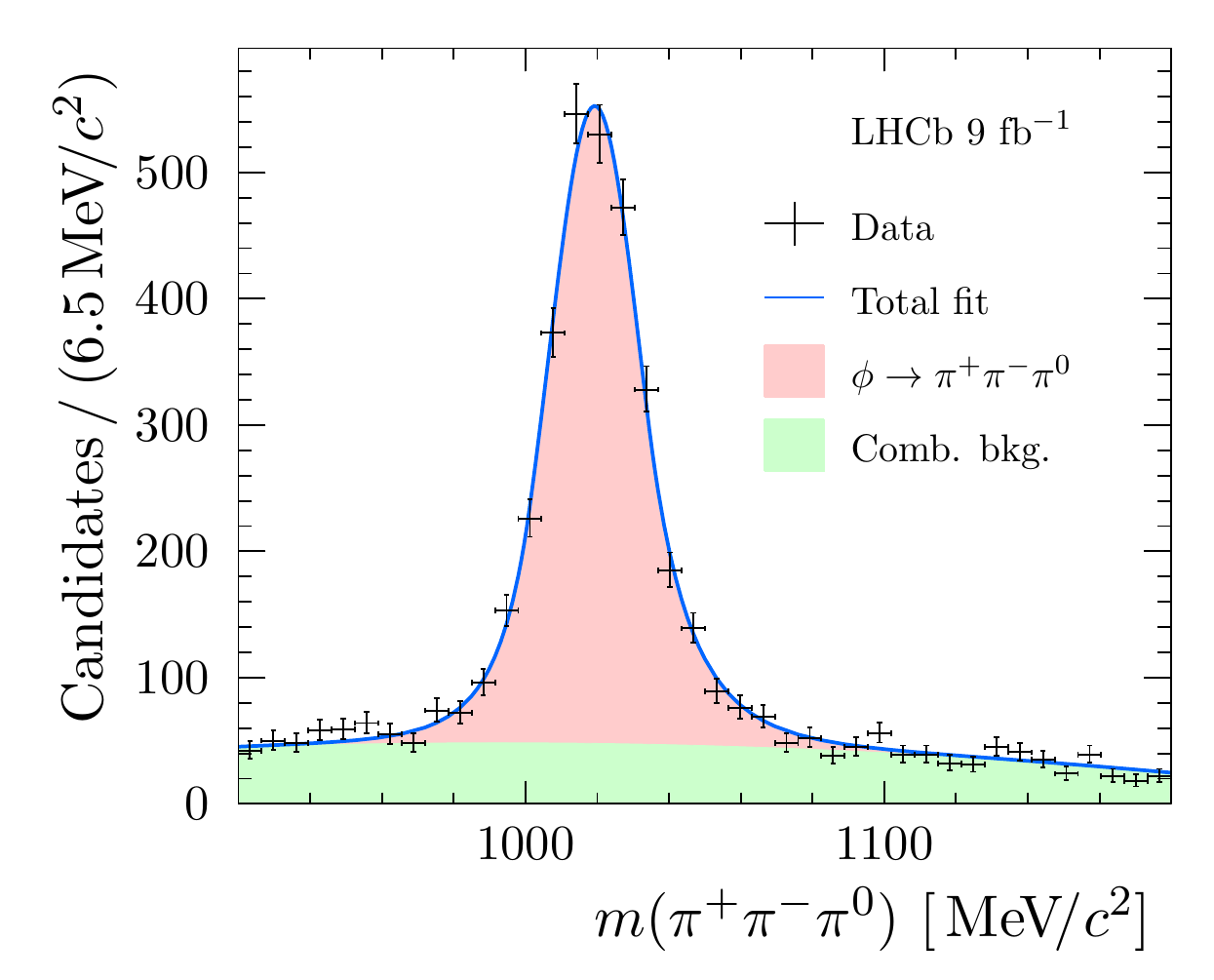}
    \caption{Mass distributions of (top) \etaz, (middle) \etapr and (bottom) $\phiz$ candidates in data. To enhance the purity, only \etazpr (\phiz) candidates with a $\jpsi\etazpr$ ($\jpsi\phiz$) mass close to the known \Bs mass are considered. The fit results show the stacked contributions from signal (in red), combinatorial background (in green) and partially reconstructed backgrounds (in gray).}
\label{fig:lineshapes}
\end{figure}

The \Bds efficiencies are corrected for small imperfections in the modelling of the \etazpr lineshapes, in particular of their peak positions and widths.
These corrections are derived from fits to the corresponding mass distributions in data, where only signal candidates in the vicinity of the \Bs peak are considered. 
Fits to the $\pip\pim\g$, $\pip\pim\piz$ and $\pip\pim\etaz$ mass distributions in data are shown in Fig.\,\ref{fig:lineshapes}, where the signal and partially reconstructed background components are modelled using simulation (the signal peak position and width are free to vary in the data fits) and combinatorial background is modelled using same-sign data.
The efficiencies of the mass requirements in data are then calculated as the integrals of the signal probability density functions in the relevant mass intervals.

The lineshape corrections vary between 3 and 5\% across the signal modes and partly cancel in the $\etaz/\etapr$ efficiency ratios, which are affected at the level of 1--3\%.
A second set of efficiency corrections addresses the modelling of diphoton lineshapes. In that case, samples of $\Bu\to\jpsi\Kstarp(\to\Kp\piz(\to\g\g))$ and $\Bs\to\jpsi\etaz(\to\g\g)$ decays are used to constrain the diphoton lineshape parameters in signal data, which allows to determine the corrections with relatively small uncertainties.\footnote{Throughout the paper, the $K^{*}(892)$ resonances are denoted as \Kstar.}
Diphoton corrections are at the level of 2\%, which drops to a negligible level when the $\etaz/\etapr$ efficiency ratios are considered.

Further efficiency corrections account for the difference between the \Bs lifetime value used in the simulation, $\tau_{\rm sim}$, and the known value, $\tau$, by assigning each simulated \Bs candidate the weight
\begin{equation}
w = \frac{\tau_{\rm sim}}{\tau} \exp\biggl(-t \biggl( \frac{1}{\tau} - \frac{1}{\tau_{\rm sim}} \biggr)\biggr),
\end{equation}
where $t$ is the true \Bs decay time.
Given the small \CP violation in the \Bs system and the \CP-even nature of the signal final states, the approximation $\tau \approx \tau_{\rm L}$ is used, where $\tau_{\rm L}=1.429 \pm 0.006\ps$ corresponds to the known lifetime of the approximately \CP-even (light) \Bs mass eigenstate~\cite{PDG2024}.
On the other hand, both the light and heavy eigenstates decay to the $\jpsi\phiz$ final state, in which case, the effective lifetime of $1.487 \ps$ is used. This value is based on the mixing and amplitude parameters from Ref.~\cite{LHCb-PAPER-2019-013} and enters the determination of the normalisation branching fraction, as detailed in Ref.~\cite{LHCb-PAPER-2020-046}.
Lifetime corrections amount to 8--9\% and 2--3\% in the signal and normalisation modes, respectively.

Additional corrections for the mismodelling of the pion PID response and \Bds kinematics largely cancel in the efficiency ratios and are therefore not applied. The associated uncertainties are discussed in Sec.~\ref{sec:syst}.

\section{Determination of the yields}
\label{sec:yields}

\subsection{Fit model}
\label{sec:yields_fit}

The \Bds yields are determined using unbinned extended maximum-likelihood fits to the $\jpsi\pip\pim\g(\g)$ mass distributions in the range 5150--5650\,\mevcc.
In the signal modes, the fit models account for the contributions from \Bz and \Bs decays and combinatorial background, whose yields are free to vary.
Also, signal decays where one photon is replaced by another photon from the event, later referred to as random photon backgrounds, are modelled in the fits. Their yields are constrained to those of the signal decays based on ratios obtained from simulation.
A similar model is used for the normalisation channel, except for the presence of the \Bz component which is expected to be negligible~\cite{LHCb-PAPER-2020-033}.

Signal contributions are modelled using a Gaussian function with a power-law tail on each side of the peak. All \Bs shape parameters are fixed from simulation, except the width which is scaled by a free parameter to account for the different resolution in data and simulation.
Given the smaller size of the available \Bz simulated samples, the corresponding peak value $\mu_{d}$ and width $\sigma_{d}$ parameters are expressed as a function of those of the \Bs modes as $\mu_{d} = \mu_{s} - \Delta_{\mu}$ and $\sigma_{d} = \sigma_{s} / R_{\sigma}$, where $\Delta_{\mu}$ and $R_{\sigma}$ are extracted from a simultaneous fit to the simulated \Bd and \Bs mass distributions.
The \Bs tail parameters are also used for the \Bz shapes.

Random photon components are modelled by a modified Gaussian function with exponential tails.
For each model, the \Bs parameters are determined using simulation and the signal scale factor is applied to the width.
Also in this case, the \Bz mean and width parameters depend on those of the \Bs modes through $\Delta_{\mu}$ and $R_{\sigma}$, while \Bs tail parameters are used for the \Bz shapes.
The choice of parametrisation for the combinatorial background is motivated by the mass distribution in same-sign data.
Except for the fit to the $\jpsi\etapr(\to\pip\pim\g)$ mass distribution whose model is further described in Sec.~\ref{sec:yields_fit_etapg}, an exponential function is used with a slope parameter that is free to vary.
Partially reconstructed backgrounds from $B^{(0,+)}\to\jpsi\etazpr\pi^{(0,+)}$ decays, where the $\etazpr\pi$ system originates mainly from the $a_{0}(980)$ scalar resonance and the pion is not reconstructed, are expected to be quite small~\cite{Albaladejo:2016mad,Liang:2014tia,Liang:2015qva}.
Moreover, this background should contribute outside the fit region and is thus ignored. Based on the branching fraction estimates from Ref.~\cite{Li:2022pkl}, similar contributions from \Bs decays are expected to be even smaller and are also neglected.

\subsubsection{\boldmath Backgrounds in the $\jpsi\etaz(\to\pip\pim\g)$ sample}
\label{sec:yields_fit_etag}

When one photon is missed, $\Bs\to\jpsi\etaz(\to\pip\pim\piz)$ and $\Bs\to\jpsi\etapr(\to\pip\pim\etaz)$ decays can satisfy the $\jpsi\etaz(\to\pip\pim\g)$ selection criteria. Their contributions relative to the \Bs signal are estimated to be 12\% and 7\%, respectively, based on the known \etazpr branching fractions~\cite{PDG2024} and simulated selection efficiencies. The estimation of the second background contribution also requires knowledge of the branching fraction ratio between $\Bs\to\jpsi\etaz$ and $\Bs\to\jpsi\etapr$ decays, which is calculated using the \phip and \phig values from the Run~1 analysis.
The relative fractions of these two background contributions in the $\jpsi\etaz(\to\pip\pim\g)$ fit are then fixed. Their shapes are modelled by a modified Gaussian function with exponential tails whose parameters are constrained using simulation. Also in this case, the signal scale factor is applied to the width of the modified Gaussian functions.
Given their smaller branching fractions, the corresponding backgrounds from \Bz decays are neglected.
The potential background contribution from $\Bz\to\jpsi\omega(\to\pip\pim\piz)$ decays with one \piz photon not reconstructed is studied and found to be negligible.

\subsubsection{\boldmath Backgrounds in the $\jpsi\etapr(\to\pip\pim\g)$ sample}
\label{sec:yields_fit_etapg}

Due to the larger dipion mass allowed in $\Bds\to\jpsi\etapr(\to\pip\pim\g)$ decays, this sample includes additional background sources, such as $\Bs\to\jpsi\phiz(\to\pip\pim\piz)$ decays in which one photon from the \piz decay is not reconstructed.
To reduce the correlation between the yield of that background and that of combinatorial background, the former is fixed to the value measured in the normalisation sample scaled by the ratio of efficiencies in simulation for the partial and full reconstruction of the  decay. Its shape is modelled by the product of an increasing and a decreasing sigmoid function, whose inflexion points and widths are determined from simulation.
Furthermore, contributions from $B^{(0,+)}\to\jpsi K^{(0,+)} \pip\pim$ decays with a kaon not reconstructed and an additional photon from the event are modelled using a modified Gaussian function with a power-law tail on the right side of the peak. The shape is determined from simulated $\Bu\to\jpsi K_{1}(1270)^{+}$ decays where both the $\Kstarz\pip$ and $\Kp\rhoz$ intermediate states of the $K_{1}(1270)^{+}$ meson are considered, as well as direct $K_{1}(1270)^{+} \to \Kp\pip\pim$ decays.
The corresponding yield is left free to vary in the fit.
Finally, the combinatorial background is modelled by the product of a Gaussian and an exponential function, which accounts for the depletion of candidates at lower masses.
Given the presence of the aforementioned peaking backgrounds in that region, the shape parameters are fixed using a fit to same-sign data while the yield is free to vary.

\subsection{Fit results}

\begin{figure}[t]
  \centering
    \includegraphics[width=0.49\textwidth]{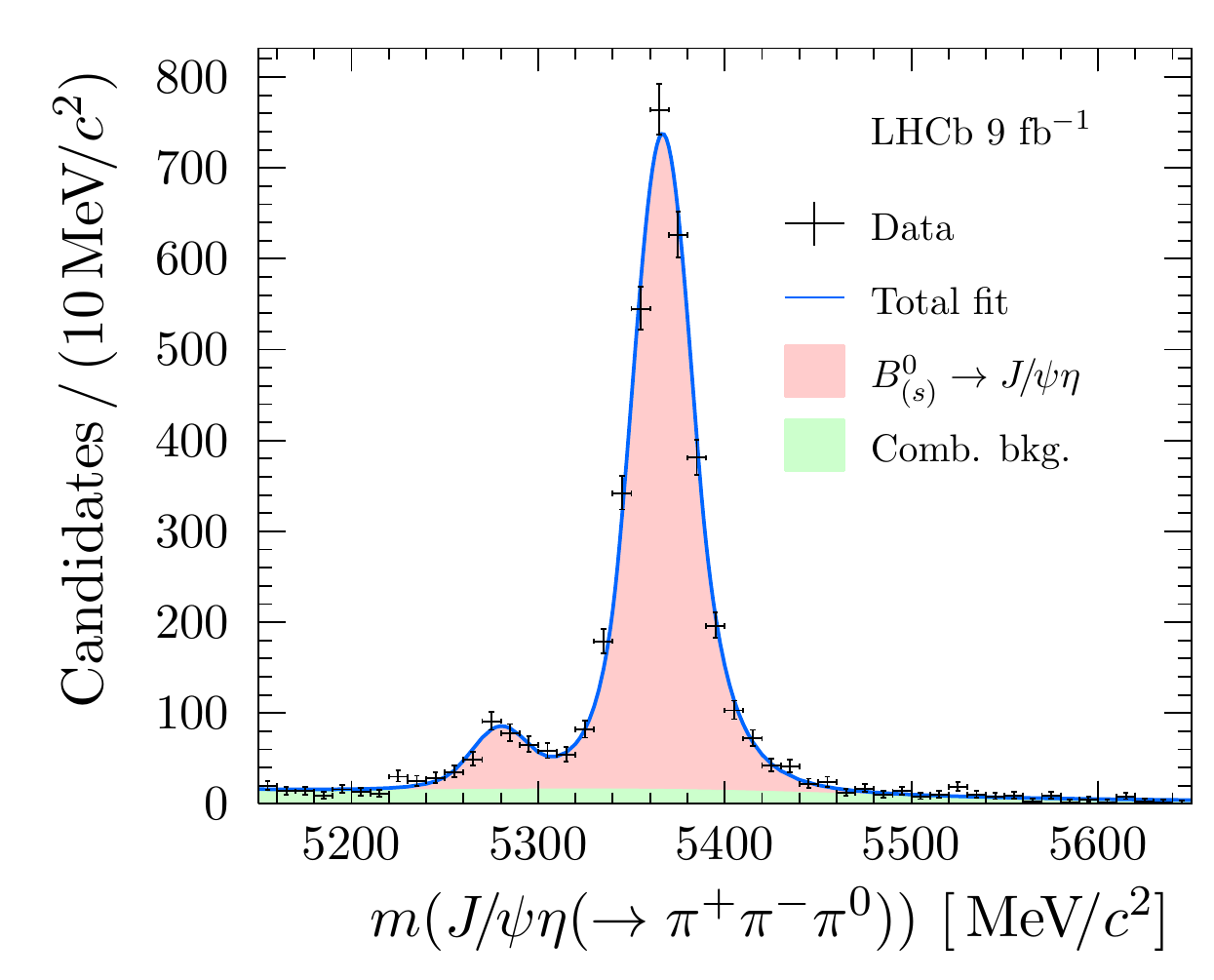}
    \includegraphics[width=0.49\textwidth]{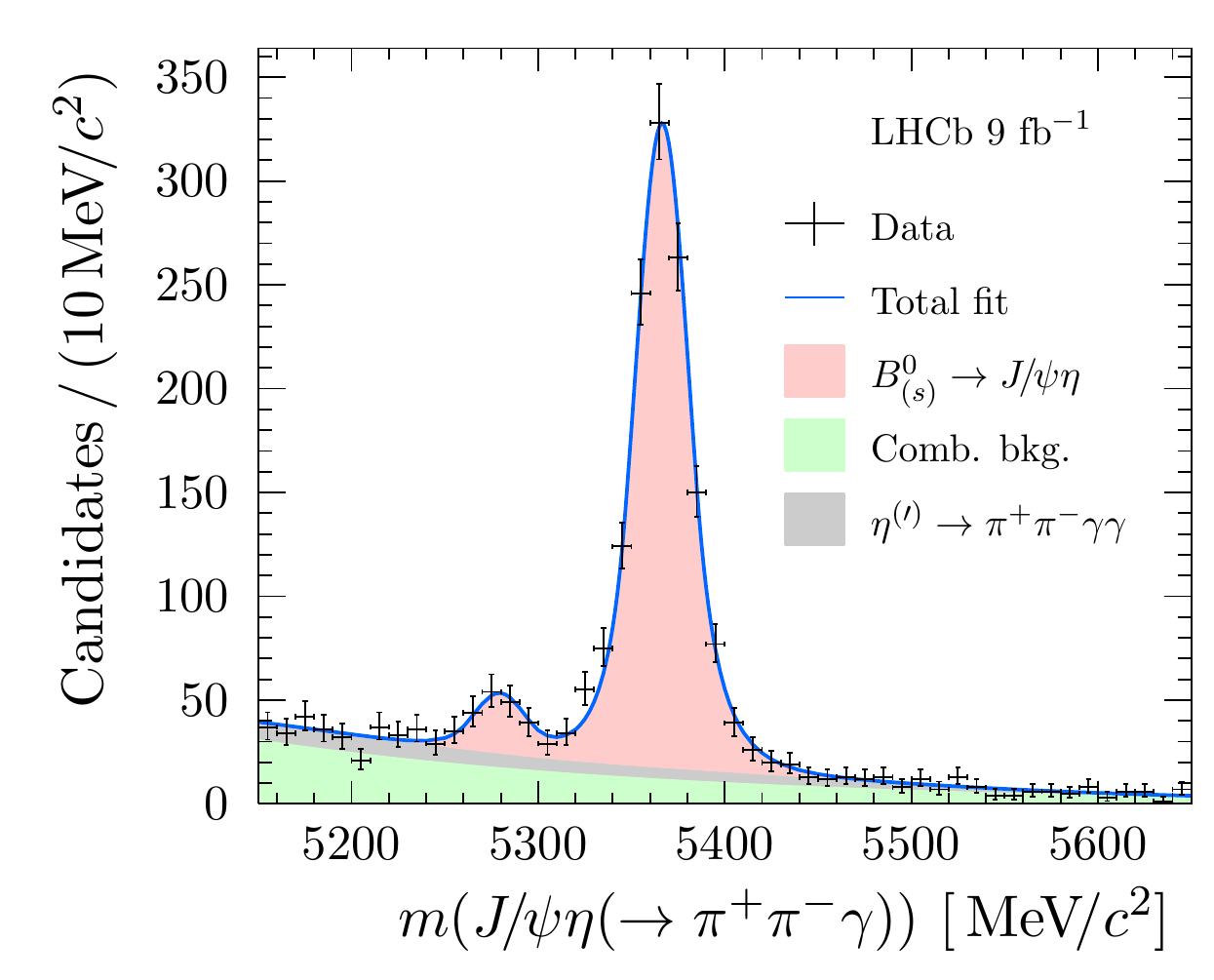}
    \includegraphics[width=0.49\textwidth]{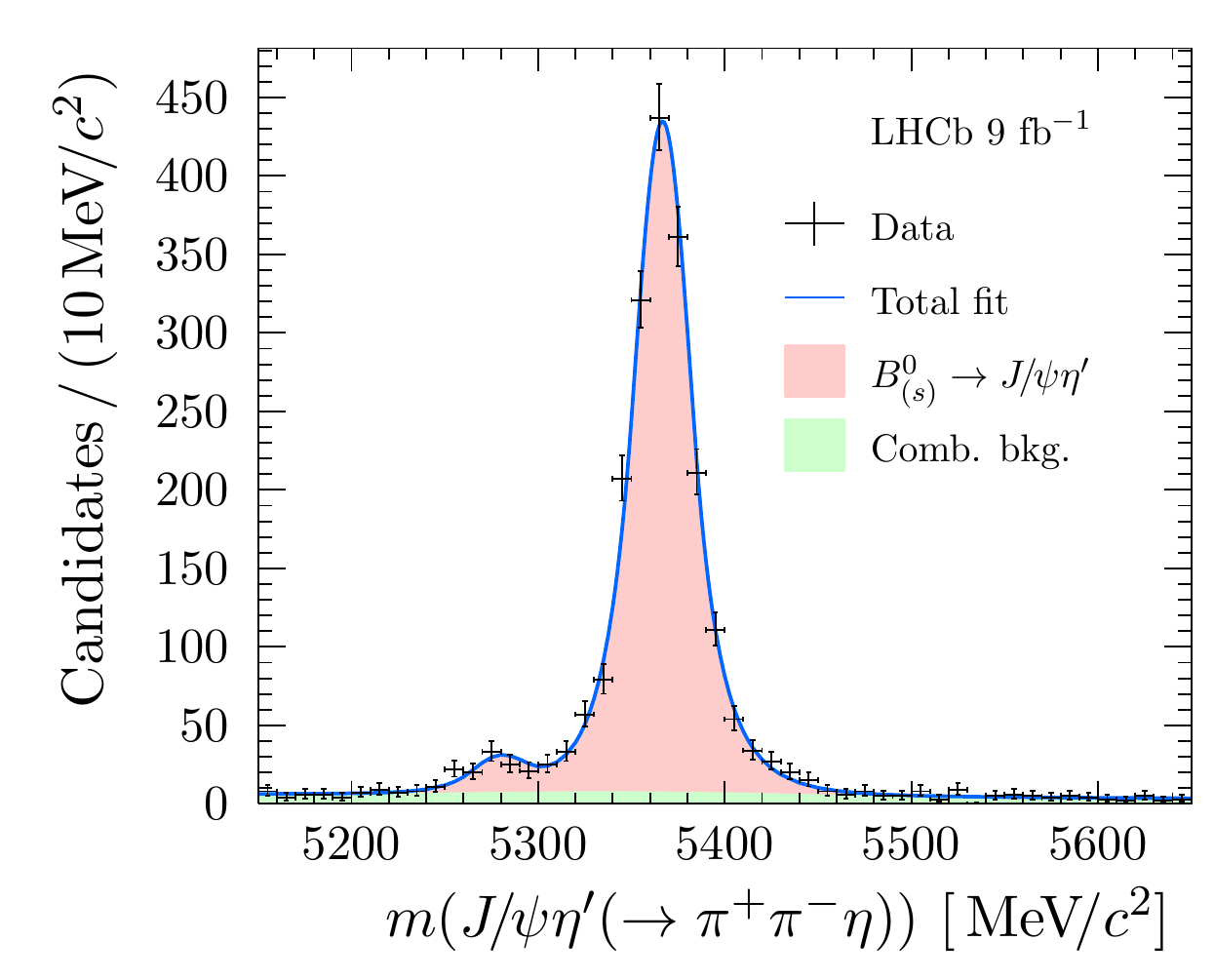}
    \includegraphics[width=0.49\textwidth]{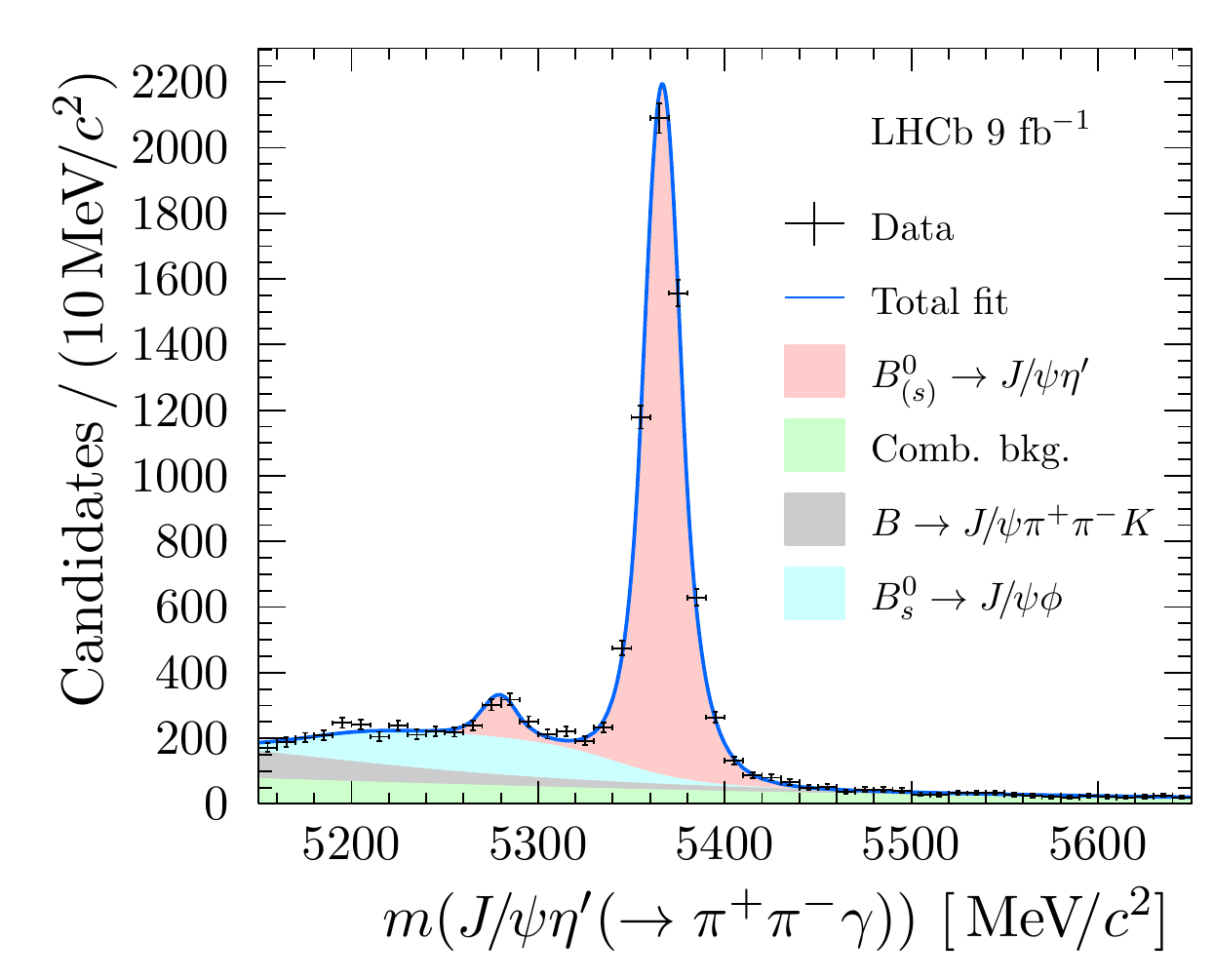}
    \includegraphics[width=0.49\textwidth]{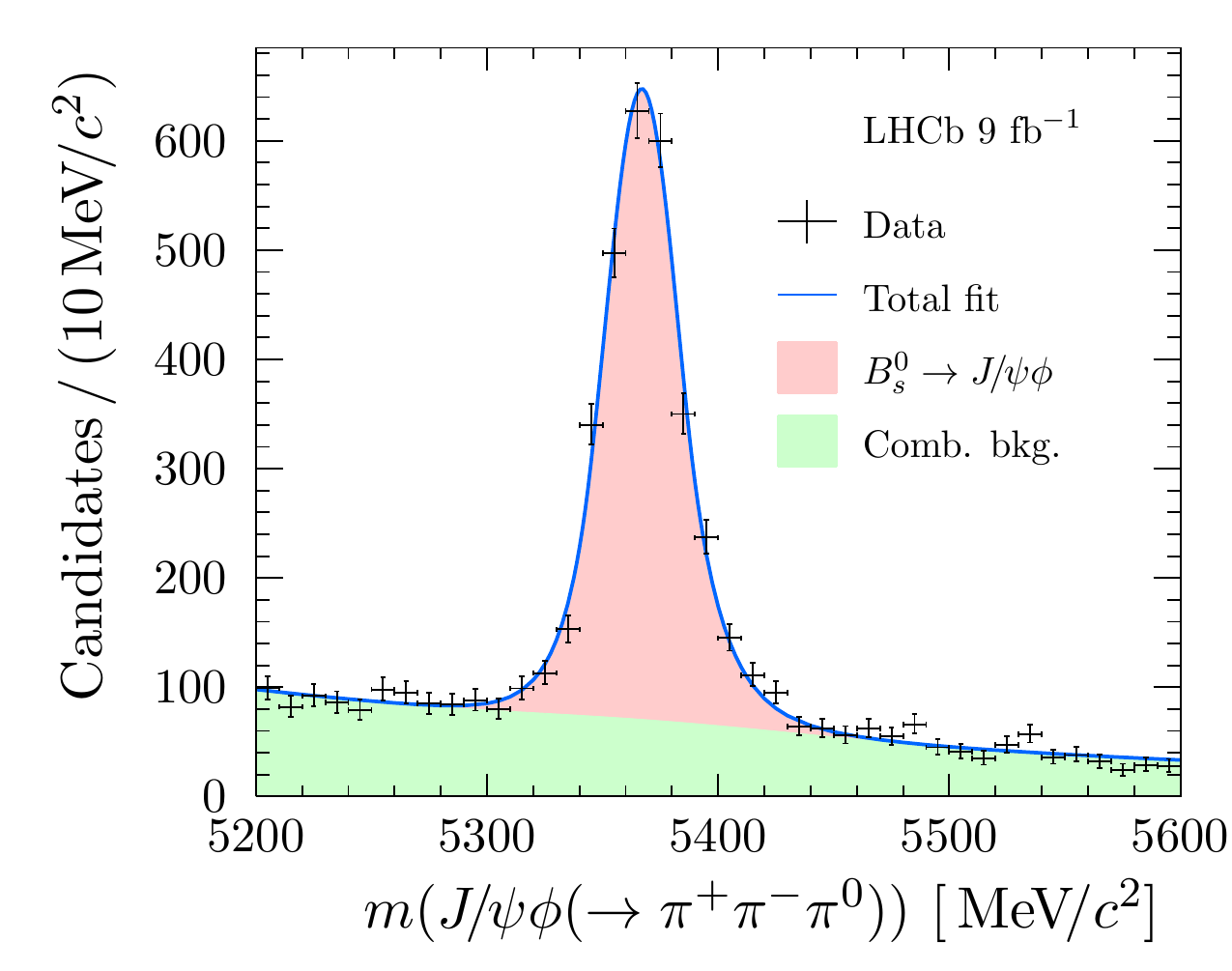}
    \caption{Mass distributions of \Bds candidates in data reconstructed in the (top) $\jpsi\etaz$, (middle) $\jpsi\etapr$ and (bottom) $\jpsi\phiz$ final states. The fit results show the stacked contributions from signal (in red), combinatorial and random-photon backgrounds (in green) and partially reconstructed backgrounds (in gray or light blue).}
\label{fig:fits}
\end{figure}

The mass distributions for the signal and normalisation modes are shown in Fig.\,\ref{fig:fits}, with fit results included.
The yields of \Bds decays in the signal and normalisation samples are listed in Table~\ref{tab:yields}. For the signal modes, a 1--5\% correlation between the \Bz and \Bs yields is observed, which is slightly more pronounced in the one-photon modes. The \Bz/\Bs yield ratios are also shown to emphasise the agreement between one-photon and two-photon results.
Statistical uncertainties at the level of 1--3\% and 8--17\% are obtained for the \Bs and \Bz modes, respectively.

\begin{table}[hbt]
\caption{Yields $N_{d,s}$ of \Bds decays in the signal and normalisation channels for different final states. The ratio and correlation coefficient ($\rho_{ds}$) between the signal yields are also indicated.}
\begin{center}\begin{tabular}{lccccc}
\hline
    & \multicolumn{5}{c}{Decay mode}\\
    & \multicolumn{2}{c}{$\Bds\to\jpsi\etaz$} & \multicolumn{2}{c}{$\Bds\to\jpsi\etapr$} & $\Bs\to\jpsi\phiz$\\
    & $\etaz\to\pip\pim\piz$ & $\etaz\to\pip\pim\gamma$  & $\etapr\to\pip\pim\etaz$ & $\etapr\to\pip\pim\gamma$ & $\phiz\to\pip\pim\piz$\\
\hline
$N_{d}$                   & \phantom{0.}285 $\pm$ \phantom{.}23 & \phantom{0.}110 $\pm$ \phantom{.}18 & \phantom{00.}94 $\pm$ \phantom{.}14 & \phantom{0.}355 $\pm$ \phantom{.}38 & \phantom{00}--     \\
$N_{s}$                   & \phantom{.}3280 $\pm$ \phantom{.}60 & \phantom{.}1244 $\pm$ \phantom{.}40 & \phantom{.}1885 $\pm$ \phantom{.}45 & \phantom{.}6002 $\pm$ \phantom{.}85 & 2572 $\pm$ 62   \\
$N_{d}/N_{s}$ [$10^{-2}$] & \phantom{00}8.7 $\pm$ 0.7           & \phantom{00}8.9 $\pm$ 1.5           & \phantom{00}5.0 $\pm$ 0.8           & \phantom{00}5.9 $\pm$ 0.7           & \phantom{00}-- \\
$\rho_{ds}$ [$10^{-2}$]   &  $-2.3$                             & $-5.1$                              & $-1.5$                              & $\phantom{-}4.6$                    & \phantom{00}--  \\ 
\hline
\end{tabular}\end{center}
\label{tab:yields}
\end{table}

\section{Systematic uncertainties}
\label{sec:syst}

The main sources of systematic uncertainties on the branching fraction ratios are summarised in Table\,\ref{tab:syst}. They pertain to the determination of the \Bds yields (Sec.~\ref{sec:syst_yield}), the evaluation of the efficiencies and their associated corrections (Sec.~\ref{sec:syst_eff}), and the precision of the various inputs that enter the calculation of the ratios (Sec.~\ref{sec:syst_ext}).

\begin{table}[t]
  \caption{Summary of relative uncertainties affecting the branching fraction ratios, in percent. In the two last columns, $R_{\etazpr,\phiz}^{\g\g}$ are the ratios between \Bs signal and normalisation modes, determined using the final states with two photons.}
  \begin{center}\begin{tabular}{lcccccccccc}
      \hline
      Source            & $R_{d}^{\g\g}$    &   $R_{d}^{\g}$    &$R_{s}^{\g\g}$ & $R_{s}^{\g}$  &$R_{\etaz}^{\g\g}$&$R_{\etaz}^{\g}$&$R_{\etapr}^{\g\g}$&$R_{\etapr}^{\g}$ & $R_{\etaz,\phiz}^{\g\g}$ & $R_{\etapr,\phiz}^{\g\g}$ \\
      \hline
      Fit model         &       4.2     &       7.8     &       0.5     &       2.4     &       1.0        &       4.9      &       4.1         &       6.5      & 1.9 & 1.9\\
      Lineshape         &       1.7     &       2.4     &       1.7     &       2.4     &       --         &       --       &       --          &       --       & 1.2 & 1.6  \\
      Photon \pt        &       1.7     &       0.4     &       1.7     &       0.4     &       --         &       --       &       --          &       --       & 0.2 & 1.5 \\
      BDT response      &       2.7     &       1.5     &       2.7     &       1.5     &       --         &       --       &       --          &       --       & 0.5 & 2.6 \\
      Sim. sample size  &       2.8     &       3.2     &       1.8     &       1.7     &       1.8        &       3.0      &       2.9         &       2.2      & 1.5 & 1.8 \\
      Decay model       &       --      &       --      &       --      &       --      &       --         &       --       &       --          &       --       & 2.2 & 2.2  \\
      \hline
      Total syst. unc.  &       6.2     &       8.9     &       4.1     &       4.1     &       2.1        &       5.8      &       5.1         &       6.9      & 3.6 & 4.9 \\
      \hline
      External inputs   &       1.7     &       2.1     &       1.7     &       2.1     &       3.1        &       3.1      &       3.1         &       3.1      & 2.9 & 3.0  \\
      \hline
      Stat. unc.        &       17.0\phantom{0}    &       19.6\phantom{0}    &       3.0     &       3.5     &       8.2        &       16.8\phantom{0}     &       15.2\phantom{0}        &       10.6\phantom{0}     & 3.1 & 3.4  \\
    \hline
  \end{tabular}\end{center}
  \label{tab:syst}
\end{table}

\subsection{Fit model uncertainties}
\label{sec:syst_yield}

The systematic uncertainties affecting the signal yields through the choice of fit models are determined by changing the shape of all components and repeating the fits. For the components that are fixed relative to the signal, the corresponding fraction is also changed. The different sources are grouped as follows.

Except for the combinatorial background, all the baseline parametrisations are determined using simulation. Either these are changed to nonparametric functions, which, in the case of the signal shapes, are convolved with a Gaussian function to account for resolution effects, or new parameters are randomly generated using the baseline models and covariance matrices.

The difference between the \Bz and \Bs lineshapes, encoded in the $\Delta_{\mu}$ and $R_{\sigma}$ parameters, is accounted for by varying these parameters according to their baseline values and uncertainties.
Similarly, the normalisation of the random photon components relative to the signal are varied according to the uncertainties of the corresponding simulation efficiencies.
For the peaking backgrounds that are constrained relative to the signal, the normalisation is changed according to the uncertainties of the simulated efficiency and known branching fractions.

In the normalisation mode, the \Bs peak is described based on simulated resonant \phiz decays. Given that the Dalitz plot in data does not reveal resonant structures, a smaller sample of nonresonant $\phiz\to\pip\pim\piz$ decays is used as an alternative model, from which a systematic uncertainty is derived.
In the case of the $\jpsi\etapr(\to\pip\pim\g)$ model, the fixed $\Bs\to\jpsi\phiz(\pip\pim\piz)$ background yield is varied according to the uncertainties of the yield measured in full reconstruction. The influence of the \phiz decay model on the efficiency is also accounted for.

For the final states other than $\jpsi\etapr(\to\pip\pim\g)$, the exponential slope parameter describing the combinatorial background is randomly drawn from a Gaussian distribution. The Gaussian mean and width are set to the value and uncertainty of the slope parameter as obtained from a fit to same-sign data in a wide mass region of 4800--6200\,\mevcc.
For the $\jpsi\etapr(\to\pip\pim\g)$ final state, the baseline model is changed to an exponential function and the same procedure is applied. However, the fit to same-sign data is now performed in the baseline fit region of 5150--6150\mevcc to avoid the low-mass region where the distribution flattens.

In summary, model uncertainties mainly impact the \Bz yields through their sensitivity to the choice of background parametrisations, with values in the range 1--7\% for the different final states. Uncertainties for the \Bs yields are negligibly small in the two-photon modes and reach 1--2\% in the one-photon modes, reflecting the larger background contributions.
These are propagated to the branching fraction ratios, with values between 0.5--7.8\%, depending on which modes are combined.

\subsection{Selection efficiency uncertainties}
\label{sec:syst_eff}

Efficiencies are calculated using simulation, under the assumption that small discrepancies between simulation and data in the modelling of selection variables cancel in the studied ratios.
The corresponding uncertainties are estimated either by correcting the simulation, as done for the \etazpr lineshapes and \Bs lifetime, or by using control modes, as done for the BDT response.

Corrections to the efficiency of the \etazpr mass requirements are determined from fits to the corresponding lineshapes in data (Fig.\,\ref{fig:lineshapes}). The statistical uncertainties on the fit parameters are propagated to the correction factors by generating several alternative normalised \etazpr shapes, and calculating the efficiency as the integral of these functions in the relevant mass intervals.
The relative variations of the efficiency range from 1--2\% for each decay mode, which are added quadratically when considering the $\etapr/\etaz$ efficiency ratios, resulting in systematic uncertainties of 1.7\% and 2.4\% in the two-photon and one-photon modes, respectively. Uncertainties for the diphoton mass requirements contribute to an uncertainty at the subpercent level, where the same method is used, together with additional lineshape constraints from $\Bs\to\jpsi\etaz(\to\g\g)$ and $\Bu\to\jpsi\Kstarp(\to\Kp\piz(\to\g\g))$ samples.
This study is performed for the normalisation channel, where uncertainties on the $\phiz/\etazpr$ efficiency ratios between 1.2\% and 1.6\% are found.

Mismodelling of the BDT classifier inputs is studied in two parts.
First, the photon \pt distributions in data and simulation are compared using the $\Bu\to\jpsi\Kstarp$ samples. For each year of data taking, weights are determined in bins of photon \pt as the ratio of the normalised distributions, and applied to simulated signal candidates.
To account for the statistical uncertainty of the \Bu signal, the assigned weights are randomly drawn from a Gaussian distribution with the mean and width set to the per-bin value and uncertainty of the weights.
The influence of the photon \pt modelling is then estimated by calculating the signal BDT efficiencies using the sum of weights instead of the number of selected candidates.
The change in the $\etaz/\etapr$ efficiency ratio is equal to 1.7\% and 0.4\% in the two-photon and one-photon modes, respectively, which are assigned as systematic uncertainties.
This study is performed for the normalisation channel, yielding 0.2\% and 1.5\% for the $\etaz/\phiz$ and $\etapr/\phiz$ efficiency ratio, respectively.

Control samples are used to assign the uncertainty for the modelling of the other classifier inputs. The decay \mbox{$\Bz\to\jpsi\rhoz(\to\pip\pim)$} is used as a proxy to the signal mode $\Bds\to\jpsi\etapr(\to\pip\pim\g)$. For the other three signal modes, which have a lower dipion mass, the decay \mbox{$\Bu\to\psitwos(\to\jpsi\pip\pim)\Kp$} is used instead.
New classifiers are trained without photon \pt information, using the baseline signal and background samples, and are then applied to the control mode candidates.
The BDT efficiencies in data are then determined by fits to the \Bz or \Bu mass distributions in data, for candidates passing or failing the requirements.
The efficiency ratios between data and simulation depart from unity by $-7\%$ to $-3\%$, depending on the control mode and BDT requirement considered.
These variations partly cancel in the double ratios, for which values of 2.7\% and $-1.5\%$ are obtained in the two-photon and one-photon modes, respectively, and assigned as systematic uncertainties.
In the normalisation channel, the uncertainties are 0.5\% and 2.6\% for the $\etaz/\phiz$ and $\etapr/\phiz$ efficiency ratio, respectively.

The uncertainties related to the finite size of the simulated samples range from 0.7--1.6\% in the \Bs signal mode to 1.5--2.6\% in the \Bz modes for which fewer samples are available.
In the normalisation mode, this uncertainty contributes to 1.1\%, with an additional 2.2\% contribution from the limited number of nonresonant \phiz decays used for an alternative efficiency determination.

Other subleading uncertainties arise from the lifetime corrections, and the modelling of the \Bds kinematics and pion PID.
The uncertainties associated to the \Bs lifetime corrections in simulation are below the percent level, as verified by varying the known values of $\tau_{\rm L}$ or the effective lifetime from Ref.~\cite{LHCb-PAPER-2020-046}, and are thus neglected.
The modelling of the \Bds kinematics is judged based on the comparison of \pt and pseudorapidity distributions of $\Bz\to\jpsi\Kstarz(\to\Kp\pim)$ and $\Bs\to\jpsi\phiz(\to\Kp\Km)$ decays in data and simulation, and is also found to have a negligible effect on the results.
Similarly, the calibration of the pion PID response would change the efficiency ratios by half a percent. As a result, no correction is applied and the corresponding uncertainties are ignored.

In summary, the dominant sources of systematic uncertainties on the \rds ratios in the evaluation of efficiencies arise from mismodelling of the \etazpr lineshapes and the BDT response, and the finite size of the simulated samples. For the $R_{\etazpr}$ ratios, only the latter source is relevant.

\subsection{External input uncertainties}
\label{sec:syst_ext}

The physical quantities used to determine the \rds ratios are the branching fractions of the reconstructed \etazpr, \phiz and \piz decays~\cite{PDG2024}, and the phase-space factors defined in Eq.\,\ref{eq:intro_phsp}. The latter are a function of the masses of the \Bds, \jpsi and \etazpr mesons~\cite{PDG2024}, whose uncertainties are small enough to be ignored.
The extraction of the angle \phip from the $R_{\etazpr}$ ratios relies on the knowledge of the \Bds masses and lifetimes, the ratio of CKM elements $|V_{cd}/V_{cs}|^{2}$, and the hadronisation fraction $f_s/f_d$, which is also relevant for the determination of the absolute branching fractions.
The hadronisation of $b$ quarks to a \Bds meson is known to depend on its transverse momentum and on the $pp$ centre-of-mass energy, as measured in Ref.~\cite{LHCb-PAPER-2020-046}, which reported $f_s/f_d$ values for an average \pt of 5\,\gevc of the decaying $B$ mesons used in that measurement.
Due to the higher reconstruction thresholds for decays with photons, the average $B$-meson \pt in the signal modes is larger, about 10\,\gevc, resulting in slightly different $f_s/f_d$ values.
Updated values are determined using the \pt-parametrisations proposed in Ref.~\cite{LHCb-PAPER-2020-046} by assigning a per-event $f_s/f_d$ value to simulated signal candidates, keeping the relative uncertainties from the published values. At $\sqs = 13\tev$ for instance, the obtained averages are 3.7\% smaller than the published values.
All the aforementioned physical quantities are listed in Table\,\ref{tab:ext_inputs}.

\begin{table}[tbh]
\caption{External inputs used in the determination of the branching fraction observables.}
\begin{center}\begin{tabular}{ccc}
    \hline
    Physical quantity & Value & Reference\\
    \hline
    $\mathcal{B}(\etaz\to\gamma\gamma)$ & (39.36 $\pm$ 0.18)\% & \multirow{8}{*}{\cite{PDG2024}} \\
    $\mathcal{B}(\etaz\to\pip\pim\piz)$ & (23.02 $\pm$ 0.25)\%&\\
    $\mathcal{B}(\etaz\to\pip\pim\gamma)$ & $\phantom{0}$(4.28 $\pm$ 0.07)\%&\\
    $\mathcal{B}(\etapr\to\gamma\gamma)$ & $\phantom{0}$(2.307 $\pm$ 0.033)\%&\\
    $\mathcal{B}(\etapr\to\pip\pim\etaz)$  & (42.5 $\pm$ 0.5)\%&\\
    $\mathcal{B}(\etapr\to\pip\pim\gamma)$ & (29.5 $\pm$ 0.4)\%&\\
    $\mathcal{B}(\piz\to\gamma\gamma)$ & (98.823 $\pm$ 0.034)\%&\\
    $\mathcal{B}(\phi\to\pip\pim\piz)$ & (15.4 $\pm$ 0.4)\%&\\
    \hline
    $m(\Bs)$ [\mevcc]    & $\phantom{0}5366.93 \pm 0.10\phantom{0}$& \multirow{5}{*}{\cite{PDG2024}}\\
    $m(\Bd)$ [\mevcc]    & $\phantom{0}5279.72 \pm 0.08\phantom{0}$&\\
    $m(\jpsi)$ [\mevcc]  & $3096.900 \pm 0.006$&\\
    $m(\etaz)$ [\mevcc]  & $\phantom{0}547.862 \pm 0.017$&\\
    $m(\etapr)$ [\mevcc] & $\phantom{0}957.78 \pm 0.06$&\\
    \hline
    $\Phi^{3}(\Bz\to\jpsi\etaz)\,/\,\Phi^{3}(\Bs\to\jpsi\etaz)$ & 0.9436&  \multirow{4}{*}{}\\
    $\Phi^{3}(\Bz\to\jpsi\etapr)\,/\,\Phi^{3}(\Bs\to\jpsi\etapr)$ & 0.9255&\\
    $\Phi^{3}(\Bz\to\jpsi\etaz)\,/\,\Phi^{3}(\Bz\to\jpsi\etapr)$ & 1.2668&\\
    $\Phi^{3}(\Bs\to\jpsi\etaz)\,/\,\Phi^{3}(\Bs\to\jpsi\etapr)$ & 1.2424&\\
    \hline
    $\tau_{\rm L}~[\sec^{-1}]$ & $(1.429 \pm 0.006) \times 10^{-12}$& \cite{PDG2024}\\
    $\tau_{\rm eff}~[\sec^{-1}]$ & $ 1.487 \times 10^{-12}$& \cite{LHCb-PAPER-2020-046}\\
    \hline
    $f_s/f_d$ (7\tev) & $0.2390 \pm 0.0076$&\multirow{4}{*}{\cite{LHCb-PAPER-2020-046}}\\
    $f_s/f_d$ (8\tev) & $0.2385 \pm 0.0075$&\\
    $f_s/f_d$ (13\tev) & $0.2539 \pm 0.0079$&\\
    $f_s/f_d$ (Run~1+2) & $0.2504 \pm 0.0078$&\\
    \hline
    $V_{cd}$ & $0.22487^{+0.00024}_{-0.00021}$&\multirow{2}{*}{\cite{CKM2021}}\\
    $V_{cs}$ & $0.973521^{+0.000057}_{-0.000062}$&\\
    \hline
\end{tabular}\end{center}
\label{tab:ext_inputs}
\end{table}

\section{Results}
\label{sec:results}

\subsection{Relative branching fractions}
\label{sec:results_sig}

The obtained branching fraction ratios for signal decays of a given $B$ flavour to $\jpsi\etaz$ and $\jpsi\etapr$ mesons determined in the two-photon final states, \rdsgg, and in the one-photon final states, \rdsg, are
\begin{equation*}
\begin{split}
  R_{d}^{\g\g} &=   0.58    \pm     0.10     \pm     0.04     \pm     0.01      ,\\
  R_{d}^{\g\phantom{\g}} &=     0.67    \pm     0.13     \pm     0.06     \pm     0.01      ,\\
  R_{s}^{\g\g} &=       1.01    \pm     0.03     \pm     0.04     \pm     0.02      ,\\
  R_{s}^{\g\phantom{\g}} &= 0.97    \pm     0.03     \pm     0.03     \pm     0.02      ,
\end{split}
\label{eq:result_rds}
\end{equation*}
where the first uncertainties are statistical, the second are systematic, and the third relates to the precision of the \etazpr branching fractions.
The determinations of \rd or \rs obtained using the two different final states are compatible within one standard deviation.

Possible biases in the central values of \rds and the accuracy of their statistical uncertainty estimates are assessed by means of pseudoexperiments. Using the baseline mass fit results, $10^4$ pseudoexperiments are generated for each final state and the corresponding mass distributions are fitted with the baseline models.
The obtained sets of yields are then combined to calculate the distributions of the $R_{d,s}^{\g(\g)}$ ratios.
For most of the ratios, the biases in the central values are below 3\% of the statistical uncertainties, and the Gaussian widths of the distributions depart from one by less than 1\%, so no corrections are applied. For the $\rdg$ ratio, a bias of 11\% and slightly asymmetric uncertainties are found. In that case, the bias is ignored and symmetric uncertainties are assumed, taking the largest of the uncertainties on either sides of the $\rdg$ distribution.
Using the pseudoexperiments, the correlation between the \rd and \rs ratios is found to be small, below --1.6\% and 5.3\% in the two-photon and one-photon modes, respectively.

The branching fraction ratios between signal decays to a given $\jpsi\etazpr$ final state for different flavours of the $B$ mesons are
\begin{equation*}
\begin{split}
  R_{\etaz}^{\g\g} &= ( 2.30    \pm     0.19     \pm     0.05     \pm     0.07      ) \times 10^{-2},\\
  R_{\etaz}^{\g\phantom{\g}} &= (   2.25    \pm     0.38     \pm     0.13     \pm     0.07      ) \times 10^{-2},\\
  R_{\etapr}^{\g\g} &= (        1.31    \pm     0.20     \pm     0.07     \pm     0.04      ) \times 10^{-2},\\
  R_{\etapr}^{\g\phantom{\g}} &= (  1.56    \pm     0.17     \pm     0.11     \pm     0.05      ) \times 10^{-2},
\label{eq:result_retazpr}
\end{split}
\end{equation*}
where the third uncertainties are associated to the precision of $f_s/f_d$.
The determinations of $R_{\etaz}$ or $R_{\etapr}$ in the one-photon and two-photon modes are compatible within one standard deviation. Using the phase-space factors from Table\,\ref{tab:ext_inputs}, the following weighted average ratios of branching fractions are obtained:
\begin{equation*}
\begin{split}
\frac{\BF (\Bz\to\jpsi\etapr)}{\BF (\Bz\to\jpsi\etaz)} &= 0.48	\pm	0.06		\pm	0.02		\pm	0.01,\\
\frac{\BF (\Bs\to\jpsi\etapr)}{\BF (\Bs\to\jpsi\etaz)} &= 0.80	\pm	0.02		\pm	0.02		\pm	0.01,\\
\frac{\BF (\Bz\to\jpsi\etaz)}{\BF (\Bs\to\jpsi\etaz)} &= (2.16	\pm	0.16		\pm	0.05		\pm	0.07		) \times 10^{-2},\\
\frac{\BF (\Bz\to\jpsi\etapr)}{\BF (\Bs\to\jpsi\etapr)} &= (1.33	\pm	0.12		\pm	0.05		\pm	0.04		) \times 10^{-2},
\label{eq:result_bfrat}
\end{split}
\end{equation*}
which are compatible with previous determinations~\cite{LHCb-PAPER-2014-056} and are more precise.
The ratios of branching fractions between signal and normalisation are measured in the two-photon modes only, yielding
\begin{equation*}
\begin{split}
\frac{\BF (\Bs\to\jpsi\etaz)}{\BF (\Bs\to\jpsi\phiz)} &= (	4.50	\pm	0.14		\pm	0.16		\pm	0.13	) \times 10^{-1},\\
\frac{\BF (\Bs\to\jpsi\etapr)}{\BF (\Bs\to\jpsi\phiz)} &= (	3.70	\pm	0.13		\pm	0.18		\pm	0.11	) \times 10^{-1},
\end{split}
\label{eq:result_rnorm}
\end{equation*}
where the last uncertainties reflect the precision of the \etazpr and \phiz branching fractions.

\subsection{Absolute branching fractions}
\label{sec:results_norm}

The branching fractions of the signal \Bs modes are calculated using the $\etazpr/\phiz$ ratios from Sec.~\ref{sec:results_sig} and the known value of the normalisation branching fraction, $\mathcal{B}(\Bs\to\jpsi\phiz)$, given by $(1.018 \pm 0.032 \pm 0.037)\times 10^{-3}$~\cite{LHCb-PAPER-2020-046}, with the first uncertainty including a contribution from $f_s/f_d$. The results are
\begin{equation*}
  \begin{split}
    \mathcal{B}(\Bs\to\jpsi\etaz)               &= (    4.58    \pm     0.14     \pm     0.16     \pm     0.26    ) \times 10^{-4},  \\
    \mathcal{B}(\Bs\to\jpsi\etapr)              &= (    3.76    \pm     0.13     \pm     0.19     \pm     0.21    ) \times 10^{-4},
    \label{eq:result_abs_bs}
  \end{split}
\end{equation*}
where the last uncertainties reflect the precision of the normalisation branching fraction.
The results for the \Bz decays, obtained by multiplying the \Bs branching fractions by the $\Bz/\Bs$ ratios reported in Sec.~\ref{sec:results_sig}, are
\begin{equation*}
  \begin{split}
    \mathcal{B}(\Bz\to\jpsi\etaz)               &= (    9.92    \pm     0.79     \pm     0.42     \pm     0.46    ) \times 10^{-6},  \\
    \mathcal{B}(\Bz\to\jpsi\etapr)              &= (    5.06    \pm     0.48     \pm     0.32     \pm     0.24    ) \times 10^{-6}.
    \label{eq:result_abs_bd}
  \end{split}
\end{equation*}
In these calculations, uncertainties from the two-photon \Bs modes are counted twice but remain marginal compared to those of the \Bz results.
On the other hand, the contribution from $f_s/f_d$ is removed.

The four results are compatible with the world averages~\cite{PDG2024} and more precise by a factor of 1.3--4.0, depending on the decay mode.
The precision of the \Bz results is still limited by statistics, whereas systematic uncertainties and external inputs dominate the \Bs results.

\subsection{Mixing angles}
\label{sec:results_angles}

\begin{figure}[tb]
  \centering
  \includegraphics[width=0.49\textwidth]{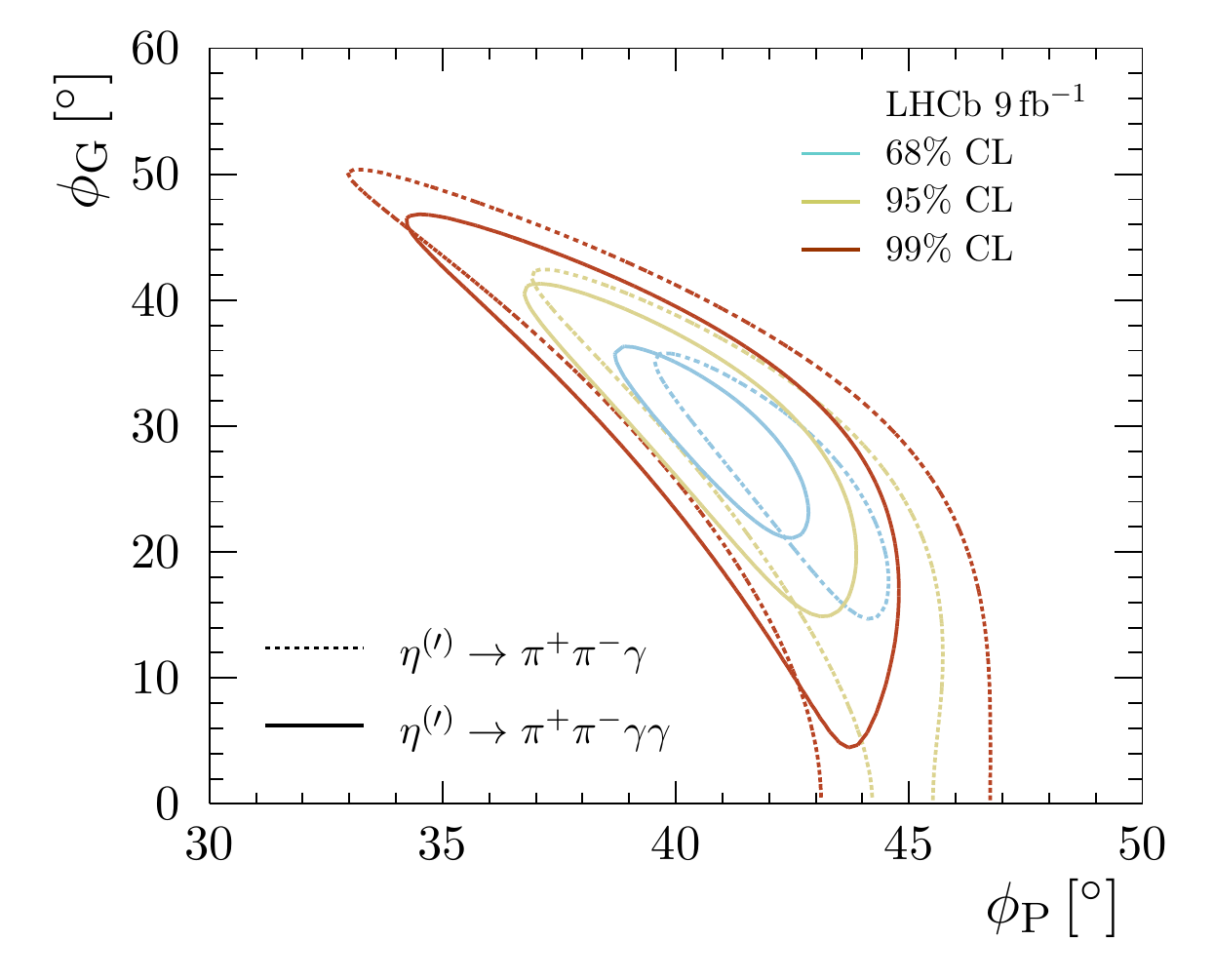}
  \includegraphics[width=0.49\textwidth]{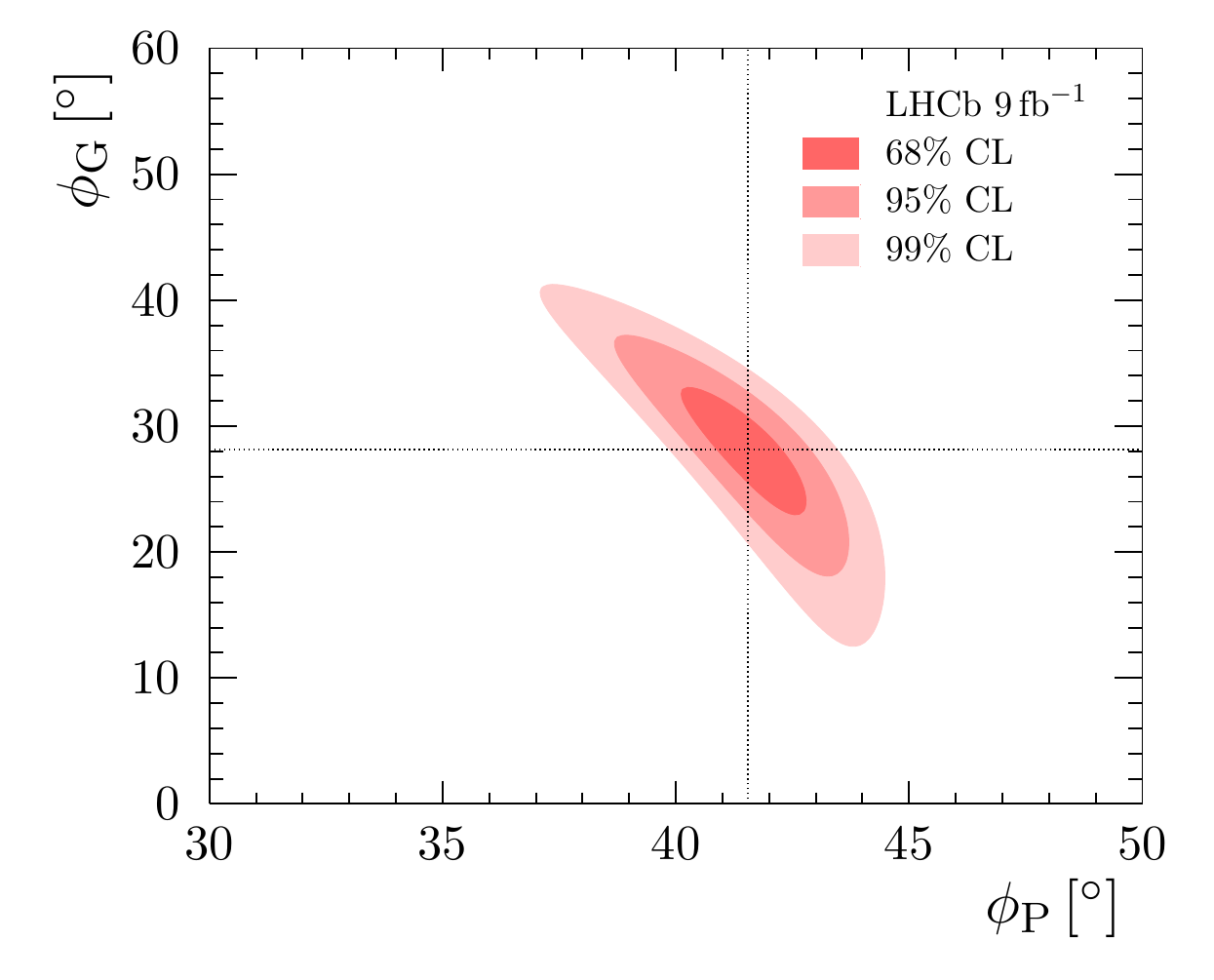}
  \caption{Contours of the two-dimensional likelihood function of \phip and \phig, corresponding to the one (68\% confidence limit, CL), two (95\% CL) and three (99\% CL) sigma regions, for (left) the one-photon and two-photon modes, and (right) the combination of these two results. The dotted lines on the right plot indicate the central values.}
  \label{fig:results_2d}
\end{figure}

Given the correlation between the \phip and \phig values determined using Eqs.~\ref{eq:intro_rd} and \ref{eq:intro_rs}, the central values and uncertainties on these observables are calculated by constructing the two-dimensional likelihood function. Similar to the Run~1 analysis~\cite{LHCb-PAPER-2014-056}, this function is defined as
\begin{equation}
  \mathcal{L} = \exp \Biggl( -\frac{1}{2} \biggl[ \biggl( \frac{\tan^{2}{\phip} \cos^{2}{\phig} - \rd}{\sigma_{\rd}} \biggr)^{2} + \biggl(  \frac{\cot^{2}{\phip} \cos^{2}{\phig} - \rs}{\sigma_{\rs}}  \biggr)^{2}  \biggr] \Biggr),
  \label{eq:results_dll}
\end{equation}
where \rds and $\sigma_{\rds}$ are the central values and total uncertainties of the ratios, respectively, as found using the one-photon or two-photon final states.
A graphical representation of the two results and their combination is shown in Fig.\,\ref{fig:results_2d}, where regions corresponding to different confidence intervals are indicated.
Numerical results for the central values and uncertainties of the angles are obtained by profiling the likelihood with respect to each of these two parameters separately for the one-photon and two-photon modes. For the combined results, the product of the two likelihood functions is used, which yields
\begin{equation*}
\begin{split}
  \phip        &= (41.6_{-1.2}^{+1.0})^{\circ},\\
  \phig        &= (28.1_{-4.0}^{+3.9})^{\circ}.
  \label{eq:results_dll_phip}
\end{split}
\end{equation*}
These two results are compatible with the Run~1 determination and supersede it.
The measured $\etaz/\etapr$ mixing angle is compatible with values found in the literature.
The gluonic angle \phig departs from zero by more than four standard deviations, as evaluated using Wilks' theorem~\cite{Wilks:1938dza}, representing the most precise measurement of this angle to date.
Besides the increased data sample and the use of a second final state to reconstruct the \etazpr mesons, the improvement in precision compared to the Run~1 result is also attributed to the fact that the uncertainty on \phig depends on its measured value. This effect can be seen in Fig.\,\ref{fig:results_2d} where the likelihood contours are stretched towards low \phig values, as expected from the corresponding smaller impact of the glueball component on the $\Bds\to\jpsi\etapr$ decay rates.

It is interesting to note that Eqs.\,\ref{eq:intro_retaz} and \ref{eq:intro_retapr} lead to different determinations of \phip, with values of $(46.9\pm0.6)^{\circ}$ and $(36.6\pm0.7)^{\circ}$, respectively. Because these equations rely on the cancellation of hadronic contributions to the amplitude of \Bz and \Bs decays to a given final-state, this discrepancy points to significant SU(3)-flavour breaking in these decays.

The angle \phip is also determined through the ratio \rz defined in Eq.\,\ref{eq:intro_r0}. Using the branching fraction of $\Bz\to\jpsi\etaz$ decays determined by this analysis and the known value for $\Bz\to\jpsi\piz$ decays~\cite{PDG2024}, the obtained value is $\rz = 0.66\pm0.08$, yielding $\phip = (35.7 \pm 4.7)^{\circ}$, which is compatible with the above \rds-based determination. As explained in Sec.~\ref{sec:intro}, a possible glueball component in the \etaz meson would decrease the ratio \rz, implying a smaller value of \phip, and is therefore disfavoured.

\section{Conclusions}
\label{sec:concl}

An updated measurement of the $\Bs\to\jpsi\etazpr$ branching fractions is presented, based on a sample of $pp$ collision data collected by the LHCb detector corresponding to an integrated luminosity of $9\invfb$.
A comparison of the branching fraction ratios to predictions from standard \etaz/\etapr mixing, which requires only one mixing angle \phip to describe the data, reveals a deficit of $\Bds\to\jpsi\etapr$ decays relative to the \etaz modes.
This result can be explained by a glueball component to the \etapr wave function and/or unexpectedly large contributions from gluon-mediated processes to the decay rates.
Parametrising the glueball component by a second angle \phig, a relatively large value of $(28.1^{+3.9}_{-4.0})^{\circ}$ is found, which differs from zero by more than four standard deviations.
An \etaz/\etazpr mixing angle \phip of $(41.6^{+1.0}_{-1.2})^{\circ}$ is determined, which is compatible and of similar precision to existing phenomenological determinations.
Absolute branching fractions are also updated using the $\Bs\to\jpsi\phiz$ channel as a normalisation. Significantly more precise values than the world averages are obtained, with factors ranging from 1.3--4.0 depending on the decay mode.
The presented results supersede those of the Run 1 analysis based on $3\invfb$ of integrated luminosity.
With the precision of the angle measurements being limited by the yields in the \Bz channels, significant improvements are expected in Run~3, based on the data from the upgraded LHCb detector.

\section*{Acknowledgements}
%
% These Acknowledgements valid from 3-May-2019
%
\noindent We express our gratitude to our colleagues in the CERN
accelerator departments for the excellent performance of the LHC. We
thank the technical and administrative staff at the LHCb
institutes.
We acknowledge support from CERN and from the national agencies:
ARC (Australia);
CAPES, CNPq, FAPERJ and FINEP (Brazil); 
MOST and NSFC (China); 
CNRS/IN2P3 (France); 
BMBF, DFG and MPG (Germany); 
INFN (Italy); 
NWO (Netherlands); 
MNiSW and NCN (Poland); 
MCID/IFA (Romania); 
%MSHE (Russia); 
MICIU and AEI (Spain);
SNSF and SER (Switzerland); 
NASU (Ukraine); 
STFC (United Kingdom); 
DOE NP and NSF (USA).
%%%%%%%%%%%%%%%%%%%%%%%%%%%%%%%%%%%%%%%%%%%%%
We acknowledge the computing resources that are provided by ARDC (Australia), 
CBPF (Brazil),
CERN, 
IHEP and LZU (China),
IN2P3 (France), 
KIT and DESY (Germany), 
INFN (Italy), 
SURF (Netherlands),
Polish WLCG (Poland),
IFIN-HH (Romania), 
%RRCKI and Yandex LLC (Russia), 
PIC (Spain), CSCS (Switzerland), 
and GridPP (United Kingdom).
%%%%%%%%%%%%%%%%%%%%%%%%%%%%%%%%%%%%%%%%%%
We are indebted to the communities behind the multiple open-source
software packages on which we depend.
%%%%%%%%%%%%%%%%%%%%%%%%%%%%%%%%%%%%%%%%%%
Individual groups or members have received support from
%ARC and ARDC (Australia); % moved to national 16/01
Key Research Program of Frontier Sciences of CAS, CAS PIFI, CAS CCEPP, 
Fundamental Research Funds for the Central Universities,  and Sci.\ \& Tech.\ Program of Guangzhou (China);
Minciencias (Colombia);
EPLANET, Marie Sk\l{}odowska-Curie Actions, ERC and NextGenerationEU (European Union);
A*MIDEX, ANR, IPhU and Labex P2IO, and R\'{e}gion Auvergne-Rh\^{o}ne-Alpes (France);
%RFBR, RSF and Yandex LLC (Russia);
Alexander-von-Humboldt Foundation (Germany);
ICSC (Italy); 
%GVA, XuntaGal, GENCAT, Inditex, InTalent and Prog.~Atracci\'on Talento, CM (Spain);
Severo Ochoa and Mar\'ia de Maeztu Units of Excellence, GVA, XuntaGal, GENCAT, InTalent-Inditex and Prog.~Atracci\'on Talento CM (Spain);
SRC (Sweden);
the Leverhulme Trust, the Royal Society and UKRI (United Kingdom).

% Do not include this in any draft (just for information in the template)
%\input{acknowledgements_intro}
% Comment this in for paper drafts; do not include this in analysis note, conference and figure reports
%\input{acknowledgements}

%\input{supplementary}
%\input{appendix}

% This should be taken out in the final paper
%\input{supplementary-app}

\addcontentsline{toc}{section}{References}
%\setboolean{inbibliography}{true}
\bibliographystyle{LHCb}
\bibliography{main,standard,LHCb-PAPER,LHCb-CONF,LHCb-DP,LHCb-TDR}

\newpage
% LHCb collaboration author list
% Data extracted on July 2nd, 2025 at 11:32am for paper reference LHCb-PAPER-2025-025
\centerline
{\large\bf LHCb collaboration}
\begin
{flushleft}
\small
R.~Aaij$^{38}$\lhcborcid{0000-0003-0533-1952},
A.S.W.~Abdelmotteleb$^{57}$\lhcborcid{0000-0001-7905-0542},
C.~Abellan~Beteta$^{51}$\lhcborcid{0009-0009-0869-6798},
F.~Abudin{\'e}n$^{57}$\lhcborcid{0000-0002-6737-3528},
T.~Ackernley$^{61}$\lhcborcid{0000-0002-5951-3498},
A. A. ~Adefisoye$^{69}$\lhcborcid{0000-0003-2448-1550},
B.~Adeva$^{47}$\lhcborcid{0000-0001-9756-3712},
M.~Adinolfi$^{55}$\lhcborcid{0000-0002-1326-1264},
P.~Adlarson$^{85}$\lhcborcid{0000-0001-6280-3851},
C.~Agapopoulou$^{14}$\lhcborcid{0000-0002-2368-0147}
% ...

{\footnotesize \it

% ...
$^{14}$Universit{\'e} Paris-Saclay, CNRS/IN2P3, IJCLab, Orsay, France\\
% ...
$^{38}$Nikhef National Institute for Subatomic Physics, Amsterdam, Netherlands\\
% ...
$^{47}$Instituto Galego de F{\'\i}sica de Altas Enerx{\'\i}as (IGFAE), Universidade de Santiago de Compostela, Santiago de Compostela, Spain\\
% ...
$^{51}$Physik-Institut, Universit{\"a}t Z{\"u}rich, Z{\"u}rich, Switzerland\\
% ...
$^{55}$H.H. Wills Physics Laboratory, University of Bristol, Bristol, United Kingdom\\
% ...
$^{57}$Department of Physics, University of Warwick, Coventry, United Kingdom\\
% ...
$^{60}$School of Physics and Astronomy, University of Glasgow, Glasgow, United Kingdom\\
$^{61}$Oliver Lodge Laboratory, University of Liverpool, Liverpool, United Kingdom\\
% ...
$^{69}$Syracuse University, Syracuse, NY, United States\\
% ...
$^{85}$Department of Physics and Astronomy, Uppsala University, Uppsala, Sweden, associated to $^{60}$\\
% ...
}
\end{flushleft}

\end{document}